\documentclass[aip,pof,amsmath,12pt, amssymb,
preprint%
]{revtex4-1}
\usepackage[utf8]{inputenc}
\usepackage{times,graphics,bm,mathptmx,graphics, psfrag,amssymb,amsmath,rotating,color,appendix,mathtools,tikz}
\usepackage{hyperref,natbib}
\usepackage{makecell}
\usepackage{mathtools,siunitx}
\usepackage{graphicx}
\usepackage{lineno,hyperref}
\usepackage{amsfonts}
\usepackage{array,booktabs}
\usepackage{multirow}
\usepackage{rotating}
\renewcommand{\arraystretch}{1.6}

\usepackage{booktabs}
\usepackage{setspace}
\usepackage{soul}
\allowdisplaybreaks
\usepackage[left=2.3cm,right=2.3cm,top=2.3cm,bottom=2.3cm]{geometry}
\hypersetup{
    colorlinks=true, 
    linktoc=all,     
    linkcolor=blue,  
}

\begin{document}
\title{Role of partial stable stratification on the onset of rotating magnetoconvection with a uniform vertical field}
\author{Tirtharaj Barman}
\author{Swarandeep Sahoo}
\email{swarandeep@iitism.ac.in}
\affiliation{Department of Applied Geophysics, Indian Institute of Technology (Indian School of Mines) Dhanbad 826004, India.}
\date{\today}

\begin{abstract}

This study examines the onset of rotating magnetoconvection under an axially imposed magnetic field in the presence of partial thermal stable stratification. Three stratification models—fully unstable, weakly stable, and strongly stable—are analyzed across Ekman numbers $E = 10^{-3}, 10^{-4}$, and $10^{-5}$ (rotation rates) and Roberts numbers $q = 0.01, 1$ and $10$ (diffusivity ratios). Magnetic back-reaction is explored by varying the Elsasser number ($\Lambda$) from 0 to 10. Symmetry-breaking effects of stable layer are assessed via an asymmetry index. Additionally, local scaling laws are derived for onset parameters and convective penetration is quantified numerically.  Results show that stable stratification promotes earlier onset and smaller-scale flows, with stronger effects in weak field regimes—hallmarks of penetrative convection. In weak magnetic fields, symmetry breaking is pronounced but weakens for strong field regime. Convective roll thickening peaks at $\Lambda = 1$, while columnarity persists in both weak and strong field regimes due to rotational constraints and elongation effects along imposed field direction, respectively. Magnetic stabilization is most effective at low to moderate values of $q$ but weakens in high $q$ regimes. Penetration depth is inversely related to the magnetic field strength and rotation rates, particularly under strong stratification, but varies non-monotonically with rotation in weakly stratified cases. In the non-magnetic limit, the critical Ekman number $E_c$, exhibiting maximum penetration effects, is obtained as $E_c = 10^{-4}$ for weak stable stratification and $E_c = 10^{-3}$ for strong stable stratification. Obtained results can provide insights into the complex interplay of various geophysical effects on planetary interiors.

\end{abstract}

\maketitle

\section{Introduction}\label{sec1}
The adiabatic temperature gradient describes the rate at which temperature changes with height (or depth) when a fluid parcel does not exchange heat with its surroundings. If the local gradient matches the adiabatic value, the parcel remains at the same temperature as the environment, experiencing zero net buoyancy—this marginal stability condition prevents the growth or decay of convective instabilities \cite{glatzmaier2013introduction, kundu2024fluid}. A gradient steeper than the adiabatic value (superadiabatic) corresponds to a thermally unstable stratification, where buoyancy forces amplify perturbations and drive convection. In contrast, a shallower gradient (subadiabatic) corresponds to a thermally stable stratification, where buoyancy suppresses convection and favors wave motions \cite{glatzmaier2013introduction}. In natural systems, however, a fluid layer can be partly unstable and partly stable, with the sequence of stable and unstable regions occurring either along the axial direction or laterally—a condition known as partial stable stratification \cite{barman2024role}. For example, in stellar interiors, a convectively unstable region is  enveloped by stable radiative zones \cite{zahn1991convective}. Similarly, the metallic hydrogen cores of Jupiter \cite{moore2022dynamo} and Saturn \cite{yan2021recipe} are surrounded by stably stratified layers as inferred by Juno mission\cite{stevenson2020jupiter} and Cassini grand finale\cite{fortney2023saturn, helled2024fuzzy}. Furthermore, in the context of Earth’s liquid outer core, a thermally or compositionally stably stratified layer is thought to overlie the convectively unstable regions at greater depths\cite{brodholt2017composition, buffett2016evidence}. Nevertheless, it remains unclear how thermally stable stratified layers can account for certain distinctive geophysical features inferred from geomagnetic field observations. Since these phenomena are closely linked to thermo-fluid interactions in the outer core, the present study focuses on the role of stable thermal stratification in modifying convection in the Earth's outer core.

While stable stratification suppresses convection, convective motions from an underlying unstable layer can penetrate into the adjacent stable region, a process referred to as penetrative convection. Nearly six decades ago, Veronis (1963)\cite{veronis1963penetrative} first investigated non-rotating penetrative convection, with applications in geophysical and astrophysical fluid dynamics. His study considered upward penetrative convection, in which an unstable layer underlies a stable layer, arising from anomalous expansion effects in ice–water experiments \cite{matsumoto2009does}. In parallel, various geophysical studies, including seismology \cite{kaneshima2018array} and geomagnetism \cite{buffett2014geomagnetic}, have inferred the presence of a stably stratified layer near the top of Earth’s outer core, overlying the convective core—an arrangement analogous to the upward penetrative convection configuration \cite{barman2024role}. The opposite arrangement of downward penetrative convection occurs when a stable layer lies beneath an unstable one, and has been studied in the contexts of stellar convection \cite{masada2013effects} and the solar dynamo \cite{hurlburt1994penetration}. Hence, focusing on the Earth’s core, we consider the upward penetrative convection configuration to investigate the role of partial stable stratification on the onset of convection.

The Earth and other planetary bodies exhibit rotation, as do stellar objects. Some stellar bodies, such as the Sun, rotate relatively slowly, whereas planetary bodies like Earth and gas giants such as Jupiter rotate rapidly. Penetrative convection behaves differently in slowly and rapidly rotating systems because inertial forces are less effective in rapidly rotating objects than in slowly rotating stellar bodies \cite{schubert2000dynamics}, leading to distinct characteristics of convection. In slowly rotating stars, rotation drives meridional circulations \cite{brun2005simulations} and differential rotation \cite{brun2017differential}, strongly influencing material mixing and heat transport between convective and radiative zones. In rapidly rotating planets, penetrative convection can trigger vigorous motions in the stable region through teleconvection \cite{zhang2002penetrative}. In Earth’s core, stable stratification near the core–mantle boundary (CMB) acts as a low-pass filter and supports wave motion, helping to explain the periodic variation of geomagnetic secular variation \cite{buffett2016evidence}. Recent investigations on spherical shell dynamo action with stable stratification near the outer boundary\cite{mukherjee2023thermal} showed that stable layer enhances the dipolarity of the magnetic field at the CMB. 

Notably, the dynamics of rotating penetrative convection in Cartesian and spherical geometry is fundamentally different\cite{zhang2002penetrative}. However, many studies have sought to explain features of penetrative convection in spherical geometry using the $f$-plane approximation. Across low to high latitudes, rotation significantly alters its characteristics\cite{cai2020penetrative, xu2024penetrative}. 
Apart from  modifying penetrative convection, rotational effects combined with spherical geometry, produces two distinct convective zones inside and outside the tangent cylinder (TC) \cite{olson2010treatise}. The TC is an imaginary cylinder aligned with the rotation axis, touching the inner core at the equator and intersecting the CMB near $70^{\circ}$ latitude. Convection is vigorous outside the TC but strongly constrained inside it by the Taylor–Proudman effect, which suppresses vertical variations \cite{olson2010treatise}. The latter is often modeled using a rotating plane-layer approximation \cite{aujogue2015onset,  sahoo2023onset}, where gravity, rotation, and thermal gradient are aligned with the vertical ($z$) direction, representative of polar regions. In this study, we adopt this plane-layer model to investigate the combined effects of rotation and stable stratification on convective instabilities.

The Earth and other planetary bodies sustain magnetic fields generated by self-sustained dynamo action in their interiors. In this geophysical context, convection and dynamo action have been extensively studied using spherical shell models \cite{mukherjee2023thermal, gubbins2007geomagnetic}, while plane-layer models have been employed to investigate convective and magnetic induction effects \cite{aujogue2015onset, sreenivasan2017confinement}. Magnetic induction can be modeled either through self-sustained dynamo models \cite{cattaneo2006dynamo} or imposed magnetic fields \cite{jones2000onset, roberts2000onset}. The latter—known as magnetoconvection—is particularly useful for studying the back-reaction of steady, large-scale magnetic fields on convective flows, providing insights into the interplay between buoyancy, Coriolis, and Lorentz forces in various geophysical and astrophysical contexts. Notable examples include the formation of flux bundles associated with sunspots \cite{peckover1978dynamic} and the generation of induced poloidal magnetic fields by convection under imposed toroidal fields in spherical shells \cite{olson1995magnetoconvection}. Further studies \cite{olson1995magnetoconvection, olson1996magnetoconvection} examined how imposed azimuthal fields ($B_{\phi}$) interact with flows inside and outside the TC in rotating spherical shells, showing that flows outside the TC remain geostrophic even for weak imposed fields, while those inside TC become magnetostrophic in the strong-field regime. Here, we take a complementary approach, employing a plane-layer setup with partial stable stratification to systematically study the onset of rotating upward penetrative magnetoconvection across a wide parameter space and draw implications for the dynamics inside the TC. 

Methodologically, fundamental studies on plane-layer magnetoconvection have largely focused on the onset of convection \cite{chandrasekhar2013hydrodynamic, jones2000onset, roberts2000onset}, typically using linear stability analysis (LSA) to determine instability characteristics such as the critical Rayleigh number ($Ra_c$), critical wavenumber ($k_c$), and the transition from steady to oscillatory modes. A comprehensive treatise \cite{chandrasekhar2013hydrodynamic} extensively covers non-rotating and rotating thermal convection, as well as magnetoconvection with a uniform axial magnetic field, while later work incorporating horizontally imposed magnetic fields \cite{jones2000convection, roberts2000onset} clarified the magnetic back-reaction on convective onset. Recent investigations on plane layer magnetoconvection have implemented axially\cite{sahoo2023onset, sreenivasan2024oscillatory} and horizontally\cite{sreenivasan2017confinement} varying magnetic field to explore the onset modal analysis. Beyond onset analysis, previous studies have explored heat transfer and flow structures in non-rotating and quasi-static magnetoconvection within 3D Cartesian box geometries, with magnetic fields imposed axially \cite{yan2019heat}, horizontally \cite{calkins2023numerical}, or at a tilt \cite{nicoski2022quasistatic}, under strong buoyancy forcing and weak to strong imposed magnetic fields. 

In addition to rotation and imposed magnetic fields, further, diffusivities strongly influence the onset characteristics, flow morphology, and heat transport of convection. In magnetoconvection or dynamo action, three types of diffusivity—thermal ($\kappa$), viscous ($\nu$), and magnetic ($\eta$)—are characterized by the thermal Prandtl number $Pr = \nu / \kappa$, the magnetic Prandtl number $Pm = \nu / \eta$, and the Roberts number $q = Pm / Pr = \kappa / \eta$, which quantify the relative importance of viscous, thermal, and magnetic diffusion. Early investigations showed that in the finite magnetic diffusion limit ($\nu \gg \kappa \lesssim \eta$), i.e., $q \ll 1$ or $q \sim 1$, convection remains stationary\cite{chandrasekhar2013hydrodynamic, roberts2000onset}, whereas in the small magnetic diffusion limit ($\nu \gg \kappa \gg \eta$), i.e., $q > 1$, it can become oscillatory depending on the rotation rate\cite{jones2000onset}. Additionally, the critical parameters for convective onset may respond differently to changes in magnetic field strength and rotation in low-$q$ regimes compared to high-$q$ regimes, and across fully unstable, weakly stable, and strongly stable thermal stratification models.
Building on this, the present study investigates the onset characteristics of rotating upward penetrative magnetoconvection with vertically imposed magnetic fields under weak and strong thermal stable stratification, exploring a broad parameter space spanning slow to rapid rotation, weak to strong imposed magnetic fields, and low to high diffusivity ratios, to comprehensively assess the magnetic field's back-reaction.

Apart from analyzing the onset characteristics of penetrative thermal and magnetoconvection, an equally important aspect is estimating the depth of convective penetration into the stable stratification. At onset, this depth is governed by the strength of stable stratification, rotation rate, and imposed magnetic field strength. In the strongly driven convection limit (high buoyancy forcing), inertia can also play a key role \cite{dietrich2018penetrative}, though it is negligible at onset. It is worth noting that penetrative effects are more limited for convection driving self-sustained magnetic fields \cite{gastine2020dynamo}, compared to convection under externally imposed fields.
Theoretical studies predict penetration depths in unbounded fluid domains \cite{takehiro2001penetration, barman2025penetration}, while numerical simulations show deeper penetration for weaker stable stratification \cite{barman2025penetration}. Increasing imposed magnetic field strength reduces penetration depth \cite{cai2020penetrative, xu2024penetrative}, and rotation induces a non-monotonic effect: in plane-layer models, penetration increases from the non-rotating to slow-rotation regime\cite{garai2022convective}, then decreases as rotation strengthens \cite{cai2020penetrative, xu2024penetrative}. In this study, we also quantitatively estimate  penetration effects on weak and strong stable stratifications across a wide range of imposed magnetic field strength.

The aim of this study is to investigate the influence of partial thermal stable stratification on the onset of magnetoconvection in a 2D plane layer subjected to a uniform vertical magnetic field, spanning a range of rotation rates and diffusivity regimes. Section (\ref{sec2}) outlines the mathematical formulation, numerical methods, and diagnostic measures. Section (\ref{sec3}) presents a detailed analysis of how stratification modifies onset thresholds, convective instability structures, symmetry breaking effects, and penetration depths under the combined effects of rotation, magnetic field, and diffusivity. Section (\ref{sec4}) discusses the results with previous studies, and Section (\ref{sec5}) summarizes the conclusions.

\section{Problem formulation}\label{sec2}
\subsection{Mathematical models}\label{sec2a}
\begin{figure}[htbp]
    \centering
    \includegraphics[clip, trim=0cm 22cm 0cm 0cm, width=1\textwidth]{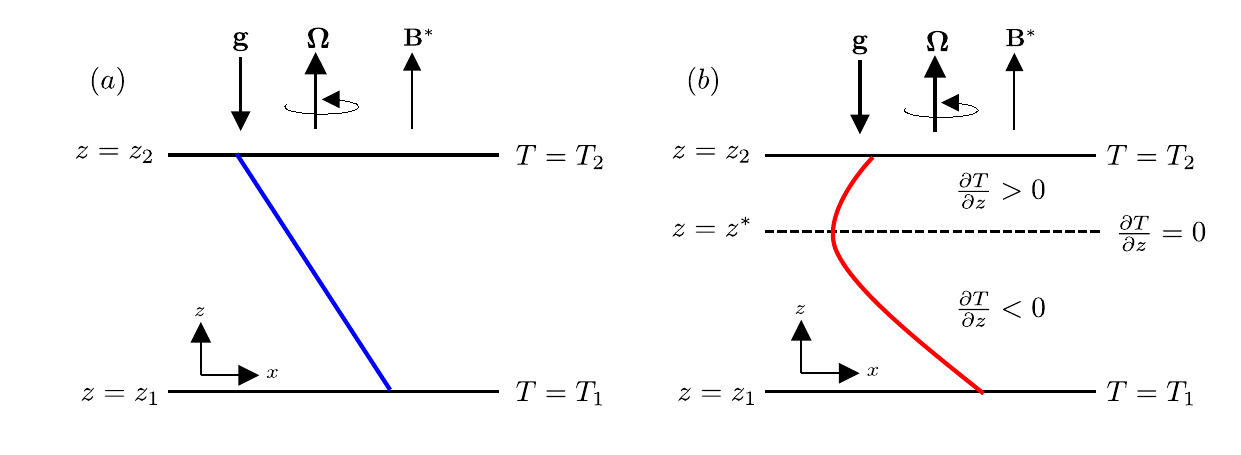}
    \caption{Schematic diagram of 2-D rotating plane layer convection model subject to background magnetic field ($\mathbf{B}^*$) in vertical direction, background rotation ($\bm{\Omega}$) directed in opposite direction of gravity ($\mathbf{g}$), and background temperature profiles, ($a$) for the reference case without thermal stable stratification (temperature profile: blue solid line), ($b$) for the case of partial stable stratification (temperature profile: red solid line).}
    \label{f_schematic}
\end{figure}

To investigate the onset of rotating magneto-convection, a fluid layer with finite thermal and electrical conductivity is considered between horizontally infinite rotating plates, as shown in figure (\ref{f_schematic}). A two-dimensional Cartesian system is chosen, where $x$ and $z$ denote horizontal and vertical directions, respectively. The plates at $z = z_1$ (bottom) and $z = z_2$ (top) are maintained at constant temperatures $T_1$ and $T_2$ ($T_1 > T_2$), creating a temperature difference $\Delta T = T_1 - T_2$ in domain $z_2 - z_1 = D$. In the reference case (figure \ref{f_schematic}a), a linear temperature profile (blue line) represents the super-adiabatic gradient across Earth’s outer core, with a hotter inner-core boundary (ICB) and cooler CMB. In the modified case (figure \ref{f_schematic}b), a stably stratified layer is introduced above $z = z^*$, where a nonlinear buoyancy profile (red line) marks the transition from thermally unstable to stable stratification. Here, the region below $z^*$ is super-adiabatic ($\partial T / \partial z < 0$), favoring convection, while the region above is sub-adiabatic ($\partial T / \partial z > 0$), suppressing convection. In both cases, rotation is applied vertically upward as $\bm{\Omega} = \Omega_0 \mathbf{\hat{z}}$, gravity acts downward as  $\mathbf{g} = -g \mathbf{\hat{z}}$, and a uniform, static magnetic field is imposed along the axial direction, as $\mathbf{B}^* = B_0 \mathbf{\hat{z}}$, to examine the back-reaction of magnetic forces on the flow. Additionally, all the fluid properties such as density($\rho$), coefficient of thermal expansion($\alpha$), viscous diffusivity($\nu$), thermal diffusivity($\kappa$), and magnetic diffusivity($\eta$) are assumed to be constant. 

Under Boussinesq approximation\cite{kundu2024fluid}, the hydromagnetic governing equations\cite{proctor1982magnetoconvection} are non- dimensionalized using $D$ as typical length scale, magnetic diffusion time, $\frac{D^2}{\eta}$, as convective time scale; $\frac{\eta}{D}$ as velocity scale, $\Delta T$ as temperature scale, and $B_0$ as magnetic field scale. Hence, the dimensionless equations are obtained as
\begin{equation} \label{eq_solenoidal_nond}
\nabla \cdot \mathbf{u} = 0 , \hspace{10pt}  \nabla \cdot \mathbf{B} = 0,
\end{equation}
\begin{equation}\label{eq_NS_nond}
\begin{split}
\frac{\partial \mathbf{u}}{\partial t} + (\mathbf{u} \cdot \nabla) \mathbf{u} + Pm E^{-1} (\mathbf{\hat{z}} \times \mathbf{u}) = - \nabla P + \Lambda Pm E^{-1} ((\nabla \times \mathbf{B})\times \mathbf{B}) + q Ra Pm T \mathbf{\hat{z}} + Pm \nabla^2 \mathbf{u},
\end{split}
\end{equation}
\begin{equation} \label{eq_temp_nond}
\frac{\partial T}{\partial t} + (\mathbf{u} \cdot \nabla) T = q \nabla^2 T + Q,
\end{equation}
\begin{equation} \label{eq_induction_nond}
\frac{\partial \mathbf{B}}{\partial t} = \nabla \times (\mathbf{u} \times \mathbf{B}) + \nabla^2 \mathbf{B},
\end{equation}
where $Q < 0$ is the dimensionless heat sink term. The dimensionless control parameters considered here are the Rayleigh number ($Ra$), thermal Prandtl number ($Pr$), magnetic Prandtl number ($Pm$), Ekman number ($E$), Elsasser number ($\Lambda$), and the Roberts number ($q$), defined as follows:
\begin{equation} \label{eq_control_pearameter}
Ra = \frac{g \alpha \Delta T D^3}{\nu \kappa}, \hspace{15pt} Pr = \frac{\nu}{\kappa}, \hspace{15pt} Pm = \frac{\nu}{\eta}, \hspace{15pt} E = \frac{\nu}{2 \Omega D^2}, \hspace{15pt} \Lambda = \frac{B_0^2}{2\Omega \rho \mu \eta }, \hspace{15pt} q  = \frac{\kappa}{\eta}.
\end{equation}
The Rayleigh number ($Ra$), representing the ratio of thermal buoyancy to viscous drag, governs the onset and intensity of convection. The thermal Prandtl number ($Pr$) compares viscous to thermal diffusion, while the magnetic Prandtl number ($Pm$) reflects the ratio of viscous to magnetic diffusion via Ohmic dissipation. Rotational influence is quantified by the Ekman number ($E$), which denotes the ratio of viscous to Coriolis forces due to background rotation. The Elsasser number ($\Lambda$) measures the relative strength of the Lorentz to Coriolis forces, indicating the influence of an imposed magnetic field. Lastly, the Roberts number ($q = \frac{Pm}{Pr}$) represents the ratio of magnetic to thermal diffusivity and is referred to as the diffusivity ratio below.

\subsection{Basic state and thermal profiles}\label{sec2b}
Convective instabilities are analyzed at onset in a plane layer driven by super-adiabatic thermal gradients within the unstable region. This is achieved by linearizing the governing equations around the reference basic state. Small perturbations in velocity, temperature, and magnetic field are introduced to evaluate their growth or decay under a given buoyancy forcing. The decomposition of all variables into basic state (denoted by $^*$) and perturbation (denoted by $^\prime$) components is expressed as
\begin{equation}
    \mathbf{u} = \mathbf{u}^* + \mathbf{u}^{\prime}, \hspace{10pt} T = T^* + T^{\prime}, \hspace{10pt} \mathbf{B} = \mathbf{B}^* + \mathbf{B}^{\prime}.
\end{equation}
All basic state variables are assumed steady. The basic velocity field is neglected ($\mathbf{u}^* = 0$) in this study, and a uniform vertical background magnetic field is imposed as $\mathbf{B}^* = \mathbf{\hat{z}}$. In addition to that, two temperature profiles are used based on thermal stratification in the basic state (denoted by $^*$) as follows: (i) fully unstable stratification, representing classical magneto-convection, and (ii) with partial stable stratification, representing penetrative magneto-convection. Here, the basic state temperature is assumed to be steady ($\partial / \partial t = 0$). Then to derive the thermal profiles the basic state temperature field ($T^*$) is assumed to vary only along the direction of gravity ($\mathbf{g}$).

\begin{figure}[htbp]
    \centering
    \includegraphics[clip, trim=2cm 14cm 1cm 0cm, width=1\textwidth]{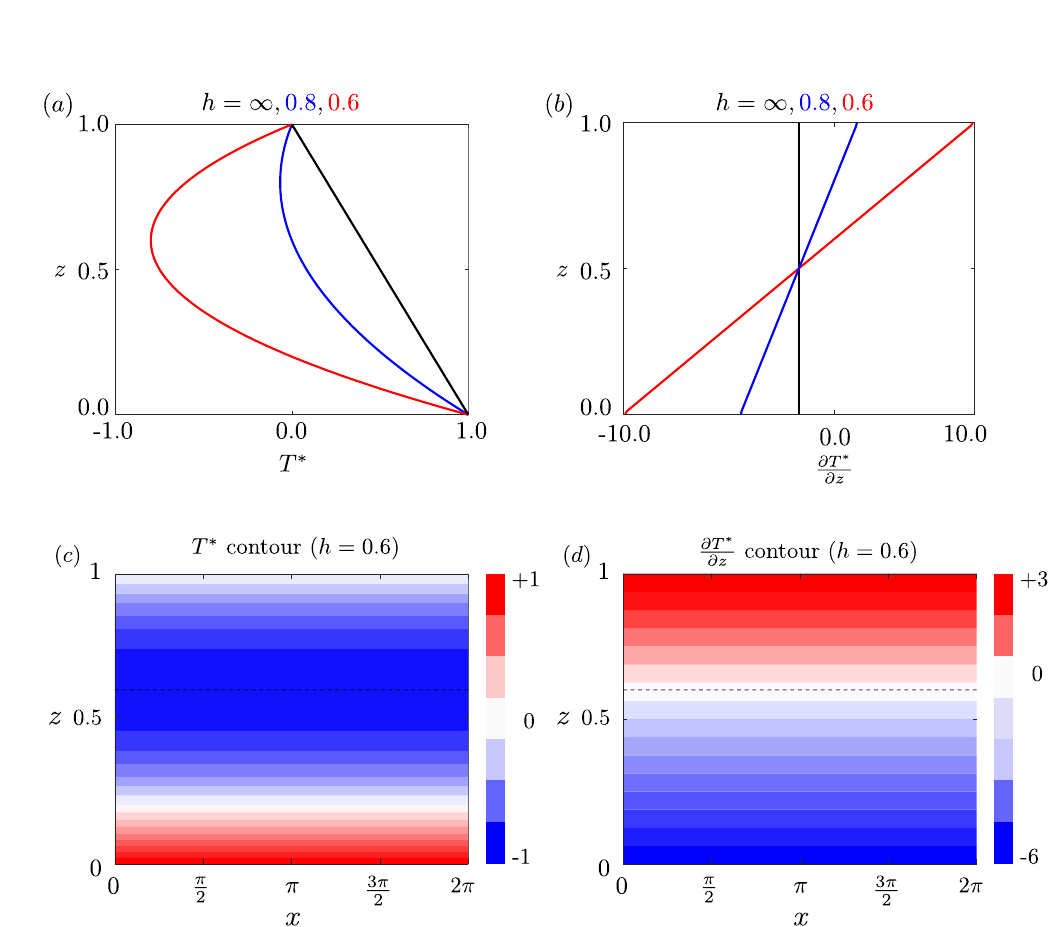}
    \caption{(a) Buoyancy profiles and (b) vertical temperature gradient profiles for three stratification cases: fully unstable stratification ($h = \infty$), weakly stable stratification ($h = 0.8$), and strongly stable stratification ($h = 0.6$). Panels (c) and (d) show the basic state temperature field $T^*(x, z)$ and its vertical gradient $\frac{\partial T^*}{\partial z}(x, z)$, respectively, for the strongly stable stratification case ($h = 0.6$). The dashed line marks the interface between the unstable and stable layers at $z = 0.6$.}
    \label{f2}
\end{figure}

Following steps in previous studies \cite{garai2022convective, barman2024role}, by setting $Q = 0$ in equation (\ref{eq_temp_nond}), the temperature profile for the reference case is obtained from basic state conduction equation as
\begin{equation}\label{eq_profile_ref}
    T^* = 1 - z,
\end{equation}
and the corresponding axial($z$) thermal gradient is estimated as
\begin{equation}\label{eq_grad_ref}
    \frac{\partial T^*}{\partial z} = -1.
\end{equation}
In the reference case, the axial temperature gradient is invariant throughout the domain and fully super-adiabatic ($\frac{\partial T^*}{\partial z} < 0$), indicating a fully unstable stratification.

The formation of a thermally stable stratification near the core–mantle boundary is primarily attributed to secular cooling \cite{lister1998stratification}. However, alternative mechanisms have also been proposed in earlier studies\cite{buffett2010stratification, gubbins2013stratified, lister2004thermal}. The same boundary conditions as in the reference case\cite{garai2022convective,barman2024role} are applied at the top and bottom boundaries. Additionally, the condition $\frac{\partial T^*}{\partial z} = 0$ is imposed at $z = h$, representing the transition height between the unstable and stable layers. The axial thermal profile with partial stable stratification is derived from basic state conduction equation with $Q$ in equation (\ref{eq_temp_nond}) following previous investigation \cite{barman2024role} as
\begin{equation} \label{eq_profile_SSL}
T^{*} =  -\frac{z^2}{1-2h} + \frac{2hz}{1-2h} + 1, 
\end{equation}
the sink term ($Q$) is derived as
\begin{equation}\label{eq_sink}
Q = \frac{2q}{1-2h},
\end{equation}
and the corresponding axial($z$) thermal gradient is obtained from equation (\ref{eq_profile_SSL}) as
\begin{equation} \label{eq_thermal_grad}
\frac{\partial T^*}{\partial z} = \frac{2 (z-h)}{2h-1}.
\end{equation}
Equation (\ref{eq_profile_SSL}) indicates that the basic state temperature varies non-linearly along the axial ($z$) direction. The corresponding thermal gradient (Equation \ref{eq_thermal_grad}) changes across the interface at $z = h$: it is negative (super-adiabatic) for $z < h$, indicating unstable stratification, and positive (sub-adiabatic) for $z > h$, indicating stable stratification. This results in two distinct axial regions, introducing a partially stable stratification. Varying the value of $h$ alters both the thickness and strength of the stable layer.

The strength of stable stratification is quantified by the Brunt–Väisälä frequency, defined as
\begin{equation} \label{eq_buoyancy_freq}
N_z = \sqrt{\alpha g \frac{\partial T^*}{\partial z}},
\end{equation}
after non-dimensionalization and normalization by $RaPr$, it becomes
\begin{equation} \label{eq_buoyancy_freq_nondim}
\frac{N_z^2}{Ra Pr} = \frac{\partial T^*}{\partial z}.
\end{equation}
This relation shows that $N_z^2 < 0$ corresponds to unstable stratification ($\frac{\partial T^*}{\partial z} < 0$), while $N_z^2 > 0$ indicates stable stratification ($\frac{\partial T^*}{\partial z} > 0$). {In the present setup, a partially stable layer exists in the region $h < z < 1$, and axial averaging using equation (\ref{eq_thermal_grad}) over this stable layer yields
\begin{equation}\label{eq_averaged_thermal_grad}
    {\langle \frac{\partial T^*}{\partial z} \rangle}_z = \int_{z=h}^{z=1} \frac{2 (z-h)}{2h-1} dz=  \frac{(1-h)^2}{(2h-1)},
\end{equation}}
which leads to the axially averaged buoyancy frequency,
\begin{equation}\label{eq_avg_buoyancy_freq}
    \frac{\langle N_z^2 \rangle}{RaPr} = \frac{(1-h)^2}{(2h -1)}
\end{equation}
a quantity that depends solely on the transition height $h$.

\begin{table}[htbp!]
\centering
\renewcommand{\arraystretch}{1}
\caption{Model parameters and definitions: $h$ denotes the transition height between unstable and stable stratification, so the stable layer has thickness $1 - h$. The reference model ($h = \infty$) has no stably stratified layer (SSL), while $h = 0.8$ and $h = 0.6$ represent weak and strong SSLs with thicknesses of 0.2 and 0.4, respectively. $Q$ is a thermal sink term used to maintain the background state. $T^*(z)$ is the prescribed axial temperature profile. $\frac{\partial T^*}{\partial z}(z)$ represents the local axial temperature gradient. $\frac{\langle N_z^2 \rangle}{RaPr}$ is the axially averaged buoyancy frequency in the stable layer ($h < z < 1$) derived from equation (\ref{eq_avg_buoyancy_freq}).}
\vspace{0.2cm}
\begin{tabular}{cccccccc}
\hline
Model &Cases & $h$& $1-h$& $\frac{Q}{q}$ & $T^*(z)$ & $\frac{\partial T^*}{\partial z}(z)$  & $\frac{\langle N_z^2 \rangle}{RaPr}$\\
\hline
1 & Without SSL & $\infty$ &0 &0 & $1 - z$  & $-1$  & 0 \\
2 & Weak SSL   & 0.8  &0.2 & - 3.33   & $1.67z^2 - 2.67z + 1$  & $10z-6$  & 2.3 \\
3 & Strong SSL & 0.6 &0.4  & -10  & $5z^2 - 6z + 1$  & $3.34z - 2.67$ & 0.0672 \\
\hline
\label{Table_models}
\end{tabular}
\end{table}

This study examines three models based on the degree of stable stratification, summarized in Table (\ref{Table_models}) where all the quantities are derived using equations (\ref{eq_profile_ref} - \ref{eq_avg_buoyancy_freq}). The first is a reference case without stable stratification, represented by $h = \infty$. The second introduces a weakly stable layer of thickness 0.2, corresponding to $h = 0.8$, while the third includes a strongly stable layer of thickness 0.4, represented by $h = 0.6$. The corresponding basic state temperature profiles ($T^*$ vs. $z$) and vertical thermal gradients ($\frac{\partial T^*}{\partial z}$ vs. $z$) are shown in Figures (\ref{f2}a) and (\ref{f2}b), respectively. The reference case ($h = \infty$, black solid line) exhibits a linear temperature profile with a constant thermal gradient along the axial direction. In contrast, both the weak ($h = 0.8$, blue solid line) and strong ($h = 0.6$, red solid line) stratification cases show non-linear temperature profiles and linearly varying thermal gradients, consistent with Equation (\ref{eq_thermal_grad}). Figures (\ref{f2}c) and (\ref{f2}d) illustrate the temperature and thermal gradient distributions for Model-2 ($h = 0.6$) across the plane layer domain. Above the transition height ($z > 0.6$), the axial thermal gradient is positive, indicating a thermally stable region, while below it ($z < 0.6$), the gradient is negative, representing an unstable region—resulting in partial stable stratification with two distinct zones. As the transition height $h$ increases, the stable layer becomes thinner, leading to a reduction in the axially averaged buoyancy frequency. Accordingly, the weakly stratified case ($h = 0.8$) exhibits a lower buoyancy frequency compared to the strongly stratified case ($h = 0.6$), as shown in Table (\ref{Table_models}).

\subsection{Perturbation equations and diagnostics}\label{sec2c}
In order to examine the influence of partial stable stratification on the onset of magnetoconvection the evolution of small perturbations to the basic state is computed using direct numerical simulations.
Substituting the decomposed variables into equations (\ref{eq_solenoidal_nond}–\ref{eq_induction_nond}) yields the dimensionless perturbation equations as
\begin{equation} \label{eq_solenoidal_pert}
\nabla \cdot \mathbf{u}^{\prime} , \hspace{10pt} \nabla \cdot \mathbf{B}^{\prime} = 0,
\end{equation}
\begin{equation}\label{eq_NS_pert}
\begin{split}
\frac{\partial \mathbf{u}^{\prime}}{\partial t} + (\mathbf{u}^{\prime} \cdot \nabla) \mathbf{u}^{\prime} + Pm E^{-1} (\mathbf{\hat{z}} \times \mathbf{u}^{\prime}) = \nabla P^{\prime} + \Lambda Pm E^{-1} ((\nabla \times \mathbf{B}^{\prime}) \times \mathbf{B}^{\prime}) \\
+ q Ra Pm T^{\prime} \mathbf{\hat{z}} + \Lambda Pm E^{-1} ((\nabla \times \mathbf{B}^{*}) \times \mathbf{B}^{\prime} + (\nabla \times \mathbf{B}^{\prime}) \times \mathbf{B}^*) + Pm \nabla^2 \mathbf{B}^{\prime},
\end{split}
\end{equation}
\begin{equation} \label{eq_temp_pert}
\frac{\partial T^{\prime}}{\partial t} + (\mathbf{u}^{\prime} \cdot \nabla)T^{\prime} + (\mathbf{u}^{\prime} \cdot \nabla)T^* = q \nabla^2 T^{\prime},
\end{equation}
\begin{equation} \label{eq_induction_pert}
\frac{\partial \mathbf{B}^{\prime}}{\partial t} = \nabla \times (\mathbf{u}^{\prime} \times \mathbf{B}^{\prime}) + \nabla \times (\mathbf{u}^{\prime} \times \mathbf{B}^*) + \nabla^2 \mathbf{B}^{\prime}.
\end{equation}
At both the top ($z_1 = 1$) and bottom ($z_2 = 0$) boundaries, no-slip conditions are applied to the velocity field, isothermal conditions to the temperature field, and pseudo-vacuum boundary conditions \cite{jones2000convection, thelen2000dynamo} to the magnetic field, given as
\begin{equation} \label{eq_boundary_condition}
(u_x^{'},u_z^{'}) = 0, \hspace{15pt} T^{'} = 0,\hspace{15pt}  (B_x^{'}, \frac{\partial B_z^{'}}{\partial z}) = 0
\end{equation}
The perturbation equations (\ref{eq_solenoidal_pert})–(\ref{eq_induction_pert}), along with the boundary conditions (\ref{eq_boundary_condition}) and basic state temperature profiles ($T^*$) for different models listed in Table (\ref{Table_models}), are solved using direct numerical simulations (DNS) with the Dedalus library \cite{burns2019dedalus}, which employs a spectral discretization approach. Chebyshev and Fourier spectral methods are used for spatial discretization in the vertical ($z$) and horizontal ($x$) directions, respectively. Depending on the resolution requirements, spatial resolutions range from $128 \times 128$ to $512 \times 512$ spectral coefficients. For time integration, a second-order Runge–Kutta method is used with a maximum time step of $5 \times 10^{-5}$ to ensure temporal accuracy.

To capture the onset characteristics, the critical Rayleigh number ($Ra_c$) is determined for a given set of control parameters ($Pr, Pm, E, \Lambda, q$), at which the perturbation fields $\mathbf{u}^{\prime}$, $\mathbf{B}^{\prime}$, and $T^{\prime}$ exhibit marginal growth. $Ra_c$ is computed iteratively by narrowing down to the minimum value with a precision of at least five significant digits. The corresponding critical horizontal wavenumber is identified as the effective inverse length scale $k_c^x$, given by
\begin{equation}\label{eq_wave_number}
    k_c = \frac{\sum \hat{u}(k_x) \cdot k_x}{\sum \hat{u}(k_x)}
\end{equation}
Here, $k_x$ represents the horizontal inverse length scale. Recent work\cite{sreenivasan2024oscillatory} investigated penetrative magnetoconvection under rapid rotation ($E = 10^{-7}, 10^{-9}$) and spatially varying magnetic fields for $q = 1$ and $10$. Similarly, using f-plane models\cite{xu2024penetrative} penetration depths are estimated at various co-latitudes ($\theta$) for $q = 1$, applying wave and convection theories under uniform magnetic fields. Extending to new regime, the present work examines the onset of rotating magnetoconvection in a plane-layer model under moderate rotation rates ($E = 10^{-3}, 10^{-4}, 10^{-5}$) with a uniform vertical magnetic field. The magnetic influence is quantified by the Elsasser number $\Lambda$ spanning from 0 to 10, where $\Lambda = 0$ denotes the non-magnetic case, $0.001 \leq \Lambda \leq 0.1$ for the weak field (rotation-dominated) regime, and $1 \leq \Lambda \leq 10$ for the strong field (magnetically dominated) regime. To explore magnetic induction across diffusivity regimes, three Roberts numbers are considered: $q = 0.01$, 1, and 10, corresponding to $(Pm, Pr) = (0.005, 0.5), (1, 1), (2, 0.2)$, respectively. The study focuses on how weak ($h = 0.8$) and strong ($h = 0.6$) partial stable stratification affect the onset, symmetry and penetration of convective instabilities.

\section{Results}\label{sec3}
In the presence of partially stable thermal stratification, the back-reaction of an axially imposed uniform magnetic field on the onset of convection is investigated. The transition from a conductive to a convective state is characterized by the overcoming of viscous resistance through buoyancy forces. This transition is influenced by the Elsasser number ($\Lambda$), the interface height between stable and unstable layers ($h$), the rotation rate ($E$), and the Roberts number ($q$). In this study, the effects of partial stratification on the spatial structure and onset characteristics of convective instabilities are first examined. Subsequently, the impacts of rotation rate ($E$) and diffusivity ratio ($q$) are analyzed, followed by a detailed assessment of penetration depth into the stable layer.

\subsection{Role of partial stable stratification}\label{sec3a}
\subsubsection{Spatial structure of convective instabilities}\label{sec3a1}

\begin{figure}[htbp]
    \centering
    \includegraphics[clip, trim=1cm 6.5cm 1cm 0cm, width=1\textwidth]{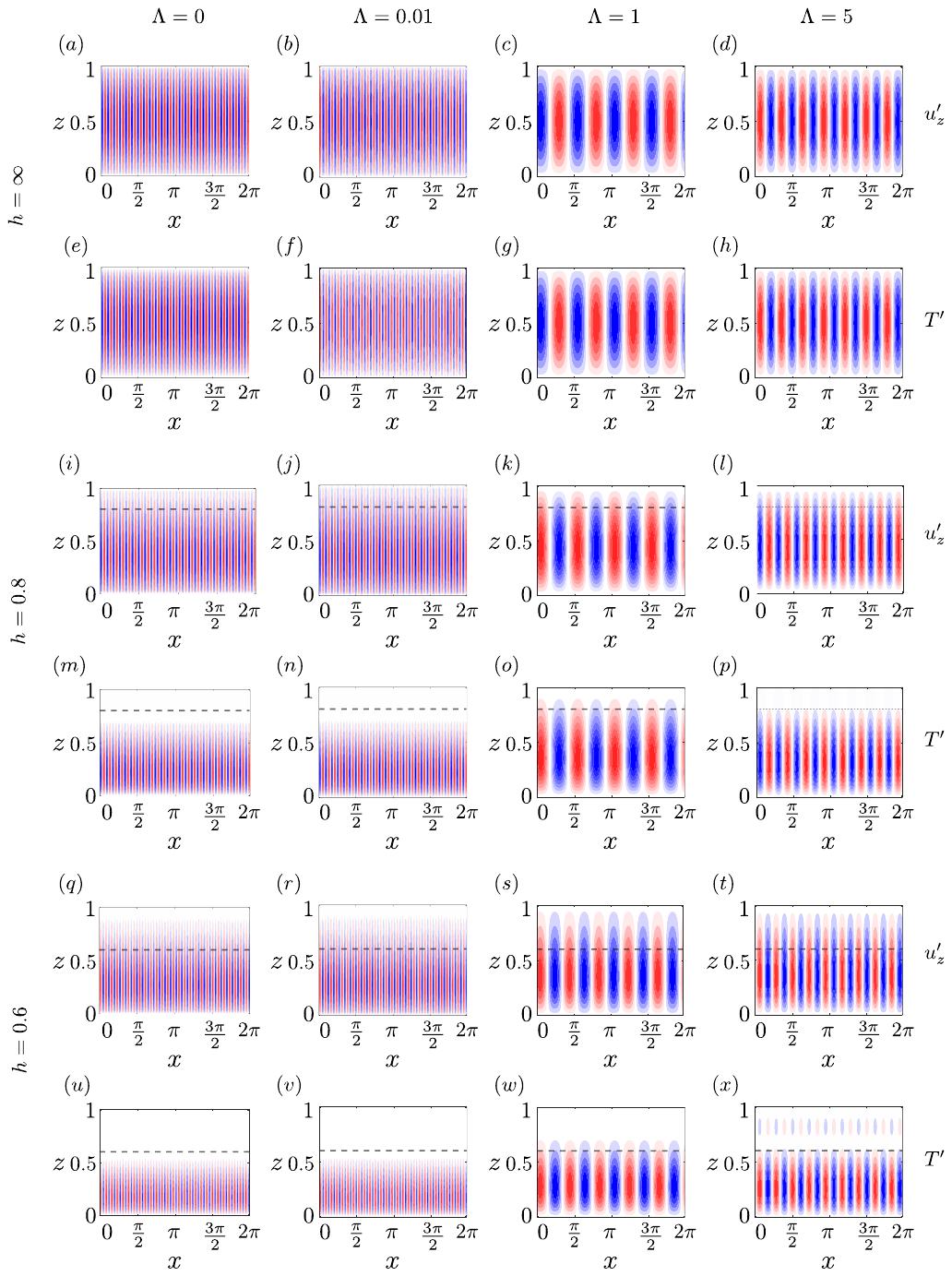}
    \caption{Axial velocity perturbation ($u_z^{\prime}$) contours (a)-(d) for $h = \infty$, (i)-(l) for $h = 0.8$, and (q)-(t) for $h = 0.6$, respectively. Temperature perturbation contours (e)-(h) for $h = \infty$, (m)-(p) for $h = 0.8$, and (u)-(x) for $h = 0.6$, respectively. The first column indicates the non-magnetic case at $\Lambda = 0$. The second, third, and fourth columns indicate increasing strength of background magnetic field as $\Lambda = 0.01, 1, 5$, respectively. Parameter regime is chosen as $E = 10^{-4}$, $q = 1 (Pr = Pm = 1)$.}
    \label{f_SSL_contour}
\end{figure}

The onset of thermal convection \cite{chandrasekhar2013hydrodynamic, barman2024role} and magnetoconvection\cite{chandrasekhar2013hydrodynamic, roberts2000onset} in a plane layer has been extensively studied to explore the fundamental properties of magnetohydrodynamics. We begin with rotating convection in the absence of a magnetic field ($\Lambda = 0$) and progressively introduce a vertical field to examine its magnetic feedback on convective onset. To illustrate the spatial structures of the vertical velocity ($u_z'$) and temperature ($T'$) perturbation at the onset (Fig. \ref{f_SSL_contour}), we fix $E = 10^{-4}$, $q = 1$, and vary $\Lambda = 0, 0.01, 1, 5$. Firstly, a fully unstable stratification case ($h = \infty$) is examined to isolate the role of weak ($h = 0.8$) and strong ($h = 0.6$) thermal stable stratification. The impact of varying rotation ($E$) and diffusivity ratio ($q$) on convective onset are illustrated in detail in sections (\ref{sec3b1}) and (\ref{sec3b2}), respectively. 

In the reference case of fully unstable stratification ($h = \infty$) and without background magnetic field ($\Lambda = 0$), axial velocity ($u_z^{\prime}$) and temperature perturbations ($T^{\prime}$) exhibit columnar structures resulting from dominance of geostrophic flow ($\partial \mathbf{u}^{\prime}/\partial z = 0$), in agreement with earlier findings \cite{garai2022convective} [Figs. \ref{f_SSL_contour}(a), \ref{f_SSL_contour}(e)]. The critical Rayleigh number ($Ra_c = 15.25925 \times 10^5$) and horizontal wavenumber ($k_c^x = 25$) for $E = 10^{-4}$ and $q = 1$, listed in Table \ref{Table_q1}, also match previous results (see Table V in reference\cite{garai2022convective}). Introducing an axial magnetic field ($\Lambda \neq 0$) modifies the convective structure. For weak fields ($\Lambda = 0.01$), the columnar rolls show negligible deviation from the non-magnetic case despite the presence of Lorentz forces (Fig. \ref{f_SSL_contour}b vs. \ref{f_SSL_contour}a), indicating minimal magnetic influence and a convection regime still controlled by rotation. Under stronger fields ($\Lambda = 1, 5$), the convective columns becomes thicker relative to the weak-field case [compare Figs. \ref{f_SSL_contour}a,b with \ref{f_SSL_contour}c,d], reflecting the relaxation of the Taylor--Proudman constraint. The flow variation follows $\partial \mathbf{u}^{\prime}/\partial z = -\Lambda \frac{\partial}{\partial z} (\nabla \times \mathbf{B}^{\prime})$ for axially imposed magnetic field, leading to quasi-2D convection that suppresses small-scale features and promotes large-scale organization \cite{mason2022magnetoconvection, sakuraba2002linear}. Interestingly, the thickening of convection rolls is non-monotonic with increasing $\Lambda$ such as the rolls are thickest at $\Lambda = 1$ [Fig. \ref{f_SSL_contour}c], where the rotational constraint and magnetic stabilization is in comparable limit. At $\Lambda = 5$, the rolls become thinner again [Fig. \ref{f_SSL_contour}d], attributed to the strong axial field exhibits flow extension in the direction of imposed field. This non-monotonic spatial behavior of convective instability persists in models with stable stratification, as shown in figures (\ref{f_SSL_contour}i–p) for thin and figures (\ref{f_SSL_contour}q–x) for thick stable stratification.

For fully unstable stratification ($h = \infty$), temperature perturbations ($T^{\prime}$) remain strongly coupled with axial velocity perturbations ($u_z^{\prime}$), as evident from the comparison of figures \ref{f_SSL_contour}(e-h) and \ref{f_SSL_contour}(a-d), respectively. Upon introducing stable stratification, the overall structural trends with increasing $\Lambda$ remain similar with $h = \infty$ model; however, $T^{\prime}$ and $u_z^{\prime}$ begin to decouple. This decoupling becomes more pronounced at higher magnetic field strengths as observed by comparing figures (\ref{f_SSL_contour}i-l) of $u_z^{\prime}$ with figure (\ref{f_SSL_contour}m-p) of $T^{\prime}$ for weak stable layer ($h = 0.8$), and the similar trend is observed for strong stable layer model as depicted in figure (\ref{f_SSL_contour}q-x). Contour plots of $T^{\prime}$ and $u_z^{\prime}$ highlight the influence of thermal stratification in both the thin ($h = 0.8$) [Figs. \ref{f_SSL_contour}(i)–\ref{f_SSL_contour}(p)] and thick ($h = 0.6$) [Figs. \ref{f_SSL_contour}(q)–\ref{f_SSL_contour}(x)] stable layer cases. Previous studies on penetrative thermal convection \cite{garai2022convective, barman2024role} reported that $T^{\prime}$ remains confined within the unstable layer due to the overlying stable stratification, while $u_z^{\prime}$ can penetrate into the stable layer. In the present study, the axially imposed background magnetic field significantly alters this behavior.
 
In the weak field regime ($\Lambda = 0, 0.01$), temperature perturbations remain confined within the unstable layer due to the stabilizing influence of the overlying stably stratified layer, irrespective of stratification strength, as shown in figures \ref{f_SSL_contour}(m, n) and \ref{f_SSL_contour}(u, v). In contrast, axial velocity penetrates into the stable region in both weak and strong stratification cases, with deeper penetration observed for weaker stratification [Figs. \ref{f_SSL_contour}(i, j) and \ref{f_SSL_contour}(q, r)], hallmark of rotating penetrative convection\cite{cai2020penetrative, barman2025penetration}. This is evident from the comparisons between figures \ref{f_SSL_contour}(i) and \ref{f_SSL_contour}(q) for $\Lambda = 0$, and \ref{f_SSL_contour}(j) and \ref{f_SSL_contour}(r) for $\Lambda = 0.01$. Thin columnar convection rolls dominate in both non-magnetic and weak-field cases, indicating strong rotational constraint. 

In the strong magnetic field regime ($\Lambda = 1, 5$), suppression of temperature perturbations by stable stratification weakens compared to the rotation-dominated case. This is evident by comparing figures \ref{f_SSL_contour}(m, n) with \ref{f_SSL_contour}(o, p) for weak stratification, and figures \ref{f_SSL_contour}(u, v) with \ref{f_SSL_contour}(w, x) for strong stratification. However, unlike the horizontally imposed magnetic field case \cite{barman2025penetration}, this weakening is non-monotonic with increasing $\Lambda$. At $\Lambda = 1$, thermal perturbations extend into the stable layer, but at $\Lambda = 5$, they are again confined to the unstable layer (compare figures (\ref{f_SSL_contour}o) to (\ref{f_SSL_contour}p) for $h = 0.8$, and figures (\ref{f_SSL_contour}w) to (\ref{f_SSL_contour}x) for $h = 0.6$). Despite this non-monotonic behavior, axial velocity continues to penetrate the stable layer for both $\Lambda = 1$ and $5$, although the extent of penetration also varies with $\Lambda$. Unlike the rotation-dominated regime ($\Lambda < 1$), where stronger stratification visibly reduces penetration (compare figures \ref{f_SSL_contour}(i, j) for $h = 0.8$ with \ref{f_SSL_contour}(q, r) for $h = 0.6$), the magnetically dominated regime ($\Lambda \geq 1$) shows no significant reduction in penetration into the stable layer with increasing stratification strength. Specifically, this is evident by comparing figures \ref{f_SSL_contour}(k, l) for weak stratification with \ref{f_SSL_contour}(s, t) for strong stratification.

Additionally, the columnar structure of convection rolls breaks down at $\Lambda = 1$, where the rolls appear thickest across all stratification cases, as seen in figures \ref{f_SSL_contour}c ($h = \infty$), \ref{f_SSL_contour}k ($h = 0.8$), and \ref{f_SSL_contour}s ($h = 0.6$). This is attributed to the combined effect of weakening axial velocity stretching along the imposed magnetic field direction and the breakdown of rotational constraints when the Lorentz and Coriolis forces become comparable at $\Lambda = 1$. Further increase in magnetic field strength to $\Lambda = 5$ partially restores columnarity across all stratification cases (figures \ref{f_SSL_contour}d, l, t). However, in this magnetically dominated regime, the partial restoration arises from the suppression of flow variations perpendicular to the magnetic field and enhanced axial stretching, rather than rotational effects as appeared in $\Lambda < 1$. In addition to changes in convective structures at onset, the threshold buoyancy force and convective length scales are also influenced by stable stratification and varying magnetic field strength, as discussed next.

\subsubsection{Axial symmetry breaking of instabilities}
\begin{figure}[htbp]
    \centering
    \includegraphics[clip, trim=1cm 17cm 1cm 0cm, width=1\textwidth]{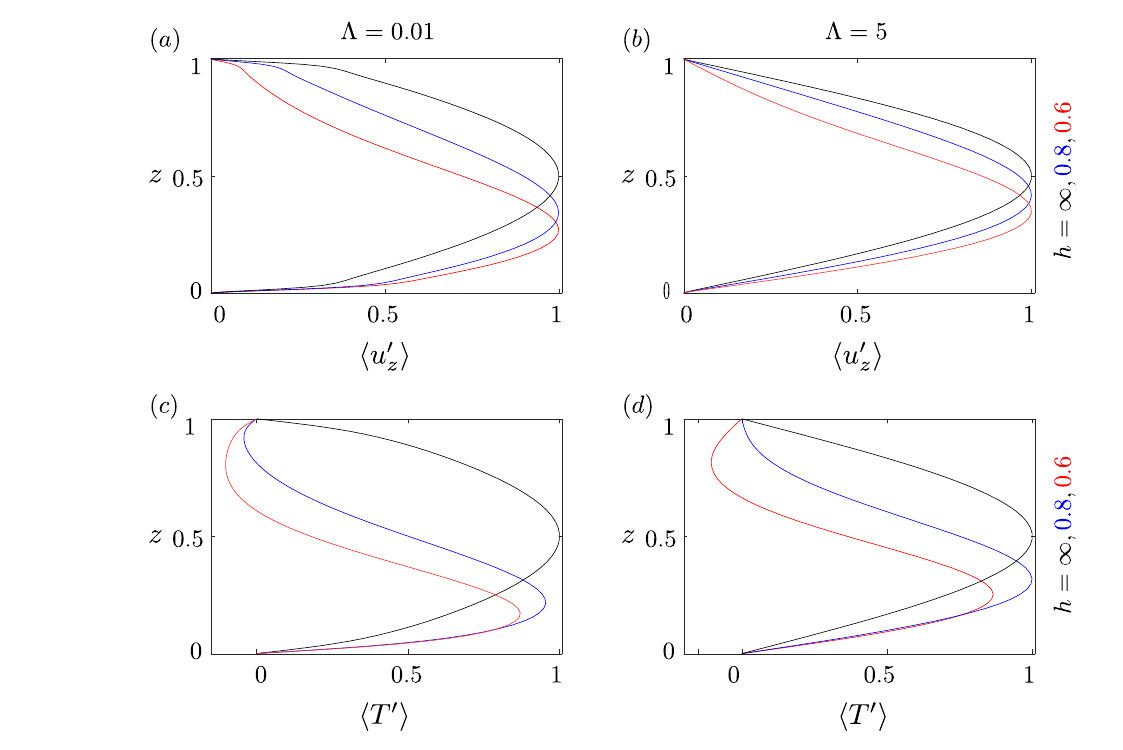}
    \caption{Axial profiles of time- and horizontally-averaged (a, b) axial velocity perturbation ($\langle \bar{u}^{\prime}_z \rangle$) and (c, d) temperature perturbation ($\langle \bar{T}^{\prime} \rangle$) for $E = 10^{-4}$ and $q = 1$. The left column (a, c) corresponds to the weak magnetic field regime ($\Lambda = 0.01$), while the right column (b, d) corresponds to the strong magnetic field regime ($\Lambda = 5$). Line colors indicate stratification strength: black — fully unstable stratification ($h = \infty$), blue — weak stable stratification ($h = 0.8$), and red — strong stable stratification ($h = 0.6$).}
    \label{f_axial_profiles_u_T}
\end{figure}

The imposition of axial magnetic field significantly alter the structure of convective instabilities. Additionally, stable stratification suppresses temperature perturbations in the unstable layer beneath the interface, while axial velocity penetrates into the stable layer. As a result, the presence of a stable layer breaks the mid-plane symmetry of convective instability. In the fully unstable case ($h = \infty$), both axial velocity and temperature perturbations remain symmetric about the mid-plane ($z = 0.5$), expressed as
\begin{equation}\label{eq_symmetry_u}
u_z^{\prime} (z > 0.5) = u_z^{\prime} (z < 0.5),
\end{equation}
\begin{equation}\label{eq_symmetry_T}
T^{\prime} (z > 0.5) = T^{\prime} (z < 0.5).
\end{equation}
This axial symmetry is preserved in the fully unstable model ($h = \infty$; black solid lines in Fig.\ref{f_axial_profiles_u_T}) across both weak and strong magnetic field regimes, as shown in figures (\ref{f_axial_profiles_u_T}a,c) for $\Lambda = 0.01$, and figures (\ref{f_axial_profiles_u_T}b,d) for $\Lambda = 5$. 

The introduction of stable stratification breaks the symmetry criterion [Equation (\ref{eq_symmetry_u} - \ref{eq_symmetry_T})] by inducing asymmetry in axial velocity and temperature perturbations. This asymmetry becomes more pronounced with increasing stratification strength, as seen in figure (\ref{f_axial_profiles_u_T})—from $h = 0.8$ (blue solid line) to $h = 0.6$ (red solid line). To quantitatively assess the relative skewness ($A_{\mu}$) of axial velocity and temperature perturbations above and below the mid-plane ($z = 0.5$), the degree of asymmetry is defined as
\begin{subequations} \label{eq_asymmetry_index}
\begin{align}
A_{u_z^{\prime}} = |\mu_{u_z^{\prime}}(z>0.5) - \mu_{u_z^{\prime}}({z<0.5})|, \\ 
A_{T^{\prime}} = |\mu_{T^{\prime}}(z>0.5) - \mu_{T^{\prime}}({z<0.5})|.
\end{align}
\end{subequations}
Here, $A_{u_z^{\prime}}$ and $A_{T^{\prime}}$ represent the relative skewness of axial velocity and temperature perturbations, respectively. Values of $A_{u_z^{\prime}} = 0$ and $A_{T^{\prime}} = 0$ indicate perfect symmetry about the mid-plane ($z = 0.5$), while $A_{u_z^{\prime}} > 0$ and $A_{T^{\prime}} > 0$ signify increasing asymmetry. Higher values correspond to stronger deviations from mid-plane symmetry. In the above definitions, skewness is evaluated separately above ($z > 0.5$) and below ($z < 0.5$) the mid-plane for both axial velocity and temperature perturbations as follows
\begin{subequations} \label{eq_skewness}
\begin{align}
\mu_{u_z^{\prime}}(z>0.5) = \frac{\bigl \langle {{u_z^{\prime}}^3(z>0.5)} \bigr \rangle}{{({u_z^{\prime}}^{rms}({z>0.5}))^3}} , 
\hspace{15pt} \mu_{u_z^{\prime}}({z<0.5}) = \frac{\bigl \langle {{u_z^{\prime}}^3({z<0.5})} \bigr \rangle}{{({u_z^{\prime}}^{rms}({z<0.5}))^3}}, \\ 
\mu_{T^{\prime}}(z>0.5) = \frac{\bigl \langle {{T^{\prime}}^3(z>0.5)} \bigr \rangle}{{({T^{\prime}}^{rms}({z>0.5}))^3}} , 
\hspace{15pt} \mu_{T^{\prime}}({z<0.5}) = \frac{\bigl \langle {{T^{\prime}}^3({z<0.5})} \bigr \rangle}{{({T^{\prime}}^{rms}({z<0.5}))^3}}.
\end{align}
\end{subequations}

For both weak ($\Lambda = 0.01$) and strong ($\Lambda = 5$) magnetic field regimes, the relative skewness values $A_{u_z^{\prime}} = 0$ and $A_{T^{\prime}} = 0$ under fully unstable stratification ($h = \infty$) indicate perfect symmetry, consistent with figure (\ref{f_axial_profiles_u_T}a–d: solid black lines). Introducing stable stratification leads to asymmetry, reflected by $A_{u_z^{\prime}} \neq 0$ and $A_{T^{\prime}} \neq 0$, with asymmetry increasing as the stratification becomes stronger, consistent with figure (\ref{f_axial_profiles_u_T}a–d: blue black lines for $h = 0.8$ and red solid line for $h = 0.6$).

In the weak field regime ($\Lambda = 0.01$), the relative skewness increases from $A_{u_z^{\prime}} = 0.22$, $A_{T^{\prime}} = 0.99$ for thin stratification ($h = 0.8$) to $A_{u_z^{\prime}} = 0.45$, $A_{T^{\prime}} = 1.74$ for thick stratification ($h = 0.6$), as seen by comparing the blue and red solid lines in figures (\ref{f_axial_profiles_u_T}a) and (\ref{f_axial_profiles_u_T}c). A similar trend is evident in the strong field regime ($\Lambda = 5$), where $A_{u_z^{\prime}}$ and $A_{T^{\prime}}$ increase from 0.14 and 0.66 ($h = 0.8$) to 1.30 and 0.77 ($h = 0.6$), respectively (Figs.\ref{f_axial_profiles_u_T}b and d). However, a comparison between weak and strong field cases shows that a stronger magnetic field reduces the degree of asymmetry. For $h = 0.8$, $A_{u_z^{\prime}}$ decreases from 0.22 at $\Lambda = 0.01$ to 0.14 at $\Lambda = 5$, and $A_{T^{\prime}}$ from 0.99 to 0.66. A similar reduction is observed for $h = 0.6$. This implies that a strong imposed magnetic field mitigates the influence of stable stratification on convective asymmetry, even at onset. This effect is likely due to the elongation of axial velocity along the magnetic field direction, which promotes columnarity in the strong field regime.  

\subsubsection{Critical values at the onset of convection}\label{sec3a2}
In addition to the modifications in spatial structure in the presence of stable stratification, the onset characteristics of magnetoconvection—specifically the critical Rayleigh number ($Ra_c$) and horizontal wavenumber ($k_c^x$)—are also modified, as illustrated in Tables \ref{Table_q0.01}, \ref{Table_q1}, and \ref{Table_q10}. The non-linear background temperature profiles, for the stable stratification, rules out a direct relationship between $Ra_c$ and transition height ($h$) of unstable-stable layer. To quantify the influence of the stable layer, the relative percentage change in $Ra_c$, denoted by $D_{Ra}$, is introduced as
\begin{equation}\label{eq_relative_drop_Ra}
D_{Ra} (\%) = \left( 1 -\frac{Ra_c^h}{Ra_c^{\infty}} \right) \times 100 \%,
\end{equation}
where $Ra_c^h$ corresponds to the critical Rayleigh number for a given stable stratification height $h$, and $Ra_c^{\infty}$ refers to the reference case without stable stratification. A higher $D_{Ra}$ value indicates a greater relative drop in $Ra_c$. This dimensionless quantity effectively captures how partial stable stratification modifies the convective onset threshold. 

\begin{table}[htbp]
\centering
\renewcommand{\arraystretch}{1}
\setlength{\tabcolsep}{12pt}
\caption{Summary of characteristic convective instability parameters for $q = 0.01$, $Pm = 0.005$, and $Pr = 0.5$. The table lists the critical Rayleigh number ($Ra_c$) and the critical horizontal wavenumber ($k_x^c$) for different rotation rates ($E$). Stratification strength is varied as follows: $h = \infty$ (fully unstable stratification), $h = 0.8$ (weak stable stratification), and $h = 0.6$ (strong stable stratification).}
\begin{tabular}{cc|cc|cc|cc}
\hline
& & \multicolumn{2}{c}{$h = \infty$}  & \multicolumn{2}{c}{$h = 0.8$ }  & \multicolumn{2}{c}{$h = 0.6$ }   \\
\hline
\multirow{11}{*}{\begin{turn}{90}$E = 10^{-3}$\end{turn}} & $\Lambda$  & $Ra_c(\times 10^4)$  & $k_x^{c}$  & $Ra_c(\times 10^4)$  & $k_x^{c}$ & $Ra_c(\times 10^4)$  & $k_x^{c}$ \\
\hline
&0	   &7.1602	&11 &6.3421  &11 &4.1987  &11\\
&0.001 &7.1521	&11 &6.3301  &11 &4.1659  &11\\
&0.01  &7.1251	&11 &6.3256  &11 &4.1498  &11\\
&0.1   &6.9987	&10 &6.2516  &11 &4.0268  &11\\ 
&1	   &5.2796	&5  &4.9019  &6  &3.3725  &7\\
&2	   &5.2412	&6  &4.9842  &6  &3.4859  &7\\
&3	   &5.9879	&7  &5.5521  &7  &3.9798  &8\\
&4     &6.8975	&7  &6.4001  &7  &4.6225  &8\\
&5	   &7.9589	&7  &7.5572  &8  &5.3112  &8\\
&10    &13.3501	&8  &12.4625 &9  &8.8612  &10\\
\hline
\multirow{11}{*}{\begin{turn}{90}$E = 10^{-4}$\end{turn}} & $\Lambda$  & $Ra_c(\times 10^5)$  & $k_x^{c}$  & $Ra_c(\times 10^5)$  & $k_x^{c}$ & $Ra_c(\times 10^5)$  & $k_x^{c}$ \\
\hline
&0	   &15.29521  &25 &13.28457  &25  &8.17601  &27\\
&0.001 &15.27091  &25 &13.21524  &25  &8.16501  &27\\
&0.01  &15.25125  &25 &13.15482  &25  &8.13811  &26\\
&0.1   &15.01986  &24 &12.90544  &25  &7.96987  &26\\
&1	   &5.89624   &4  &5.62142   &4   &3.98975  &5\\
&2	   &5.45897   &5  &5.14554   &6   &3.55128  &6\\
& 3	   &5.89547   &6  &5.50129   &7   &3.90812  &8\\
& 4    &6.65782   &7  &6.29875   &7   &4.41141  &9\\
& 5	   &7.59247   &8  &7.15001   &9   &4.91328  &10\\
& 10   &12.41051  &11 &11.45412  &11  &7.87998  &13\\
\hline
\multirow{11}{*}{\begin{turn}{90}$E = 10^{-5}$\end{turn}} & $\Lambda$  & $Ra_c(\times 10^6)$  & $k_x^{c}$  & $Ra_c(\times 10^6)$  & $k_x^{c}$ & $Ra_c(\times 10^6)$  & $k_x^{c}$ \\
\hline
&0	   &34.515671  &55 &29.525002  &57 &18.174725 &60\\
&0.001 &34.501456  &55 &29.520947  &57 &18.171052 &60\\
&0.01  &34.469854  &55 &29.499759  &57 &18.146238 &60\\
&0.1   &15.055671  &55 &29.155899  &56 &17.884957 &60\\ 
&1     &5.898598   &4  &5.758941   &4  &4.254897  &5\\
&2	   &5.720151   &5  &5.251456   &5  &3.678578  &6\\
&3	   &6.258694   &6  &5.750491   &7  &3.954658  &8\\
&4     &6.752564   &7  &6.255411   &7  &4.398569  &8\\
&5	   &7.589435   &8  &7.251456   &8  &4.412759  &9\\
&10	   &12.309975  &11 &11.212457  &12 &7.801541  &14\\
\hline
\label{Table_q0.01}
\end{tabular}
\end{table}

\begin{table}[htbp]
\centering
\renewcommand{\arraystretch}{1}
\setlength{\tabcolsep}{12pt}
\caption{Summary of characteristic convective instability parameters for $q = 1$, $Pm = 1$, and $Pr = 1$. The table lists the critical Rayleigh number ($Ra_c$) and the critical horizontal wavenumber ($k_x^c$) for different rotation rates ($E$). Stratification strength is varied as follows: $h = \infty$ (fully unstable stratification), $h = 0.8$ (weak stable stratification), and $h = 0.6$ (strong stable stratification).}
\begin{tabular}{cc|cc|cc|cc}
\hline
& & \multicolumn{2}{c}{$h = \infty$}  & \multicolumn{2}{c}{$h = 0.8$ }  & \multicolumn{2}{c}{$h = 0.6$ }   \\
\hline
\multirow{11}{*}{\begin{turn}{90}$E = 10^{-3}$\end{turn}} & $\Lambda$  & $Ra_c(\times 10^4)$  & $k_x^{c}$  & $Ra_c(\times 10^4)$  & $k_x^{c}$ & $Ra_c(\times 10^4)$  & $k_x^{c}$ \\
\hline
&0	   &7.1129 & 11  &6.3069  & 11 &4.0999  &12\\
&0.001 &7.1131 & 11  &6.3055  & 11 &4.0987  &12\\
&0.01  &7.0988 & 11  &6.2912  & 11 &4.0905  &12\\
&0.1   &6.9641 & 10  &6.1601  & 11 &4.0171  &11\\ 
&1     &5.0381 &  5  &4.6951  & 5  &3.3159  &7\\
&2	   &5.0879 &  5  &4.7352  & 6  &3.4141  &7\\
&3	   &5.8124 &  6  &5.4425  & 6  &3.9298  &7\\
&4     &6.7648 &  7  &6.2977  & 7  &4.5481  &8\\
&5	   &7.7668 &  7  &7.2475  & 7  &5.2235  &8\\
&10	   &13.1555 & 8  &12.2152 & 9  &8.7475  &10\\
\hline
\multirow{11}{*}{\begin{turn}{90}$E = 10^{-4}$\end{turn}} & $\Lambda$  & $Ra_c(\times 10^5)$  & $k_x^{c}$  & $Ra_c(\times 10^5)$  & $k_x^{c}$ & $Ra_c(\times 10^5)$  & $k_x^{c}$ \\
\hline
&0	   &15.25989   &25 &13.13501 &25 &8.15205  & 27\\
&0.001 &15.25698   &25 &13.13222 &25 &8.15005  & 27\\
&0.01  &15.23691   &25 &13.10702 &25 &8.13348  & 27\\
&0.1   &15.00279   &24 &12.86421 &25 &7.95879  & 26\\ 
&1     &5.539809   &4  &5.19939  &4  &3.74991  & 5\\
&2	   &5.108599   &5  &4.77724  &5  &3.43192  & 6\\
&3	   &5.676001   &6  &5.28909  &6  &3.77635  & 7\\
&4     &6.481052   &7  &6.01071  &7  &4.26859  & 8\\
&5	   &7.372005   &7  &6.81625  &8  &4.81587  & 9\\
&10	   &12.17855   &10 &11.20151 &11 &7.82945  &13\\
\hline
\multirow{11}{*}{\begin{turn}{90}$E = 10^{-5}$\end{turn}} & $\Lambda$  & $Ra_c(\times 10^6)$  & $k_x^{c}$  & $Ra_c(\times 10^6)$  & $k_x^{c}$ & $Ra_c(\times 10^6)$  & $k_x^{c}$ \\
\hline
& 0	   &34.502295  & 55  &29.514911 & 57 &18.167599  &60\\
&0.001 &34.498901  & 55  &29.511981 & 57 &18.164725  &60\\
&0.01  &34.466052  & 55  &29.476695 & 57 &18.139087  &60\\
&0.1   &34.110523  & 54  &29.134852 & 56 &17.800521  &59\\ 
&1     &5.703592   &  4  &5.335082  & 4  &3.886898   &5\\
&2	   &5.158921   &  5  &4.813519  & 5  &3.469598   &6\\
&3	   &5.701357   &  6  &5.300014  & 6  &3.788876   &7\\
&4     &6.491805   &  7  &6.001899  & 7  &4.265968   &8\\
&5	   &7.343225   &  7  &6.802171  & 8  &4.800419   &9\\
&10	   &12.060255  & 10  &11.077509 & 11 &7.710198   &13\\
\hline
\label{Table_q1}
\end{tabular}
\end{table}

\begin{table}[htbp]
\centering
\renewcommand{\arraystretch}{1}
\setlength{\tabcolsep}{12pt}
\caption{Summary of characteristic convective instability parameters for $q = 10$, $Pm = 2$, and $Pr = 0.2$. The table lists the critical Rayleigh number ($Ra_c$) and the critical horizontal wavenumber ($k_x^c$) for different rotation rates ($E$). Stratification strength is varied as follows: $h = \infty$ (fully unstable stratification), $h = 0.8$ (weak stable stratification), and $h = 0.6$ (strong stable stratification).}
\begin{tabular}{cc|cc|cc|cc}
\hline
& & \multicolumn{2}{c}{$h = \infty$}  & \multicolumn{2}{c}{$h = 0.8$ }  & \multicolumn{2}{c}{$h = 0.6$ }   \\
\hline
\multirow{11}{*}{\begin{turn}{90}$E = 10^{-3}$\end{turn}} & $\Lambda$  & $Ra_c(\times 10^4)$  & $k_x^{c}$  & $Ra_c(\times 10^4)$  & $k_x^{c}$ & $Ra_c(\times 10^4)$  & $k_x^{c}$ \\
\hline
& 0	   &6.0649 &9 &5.9585  &9 &3.9304  &10\\
&0.001 &6.0671 &9 &5.9645  &9 &3.9312  &10\\
&0.01  &6.0819 &9 &5.9590  &9 &3.9359  &10\\
&0.1   &6.2459 &9 &6.0440  &9 &4.0169  &11\\ 
&1     &5.0339 &6 &4.6925  &5 &3.3152  &7\\
&2	   &4.2567 &7 &4.0650  &4 &3.0475  &6\\
&3	   &3.5841 &7 &3.4725  &4 &2.7601  &5\\
&4     &3.2279 &7 &3.2200  &4 &2.6515  &5\\
&5	   &3.1375 &8 &3.1300  &4 &2.6355  &5\\
&10	   &3.2685 &8 &3.5925  &4 &3.1379  &5\\
\hline
\multirow{11}{*}{\begin{turn}{90}$E = 10^{-4}$\end{turn}} & $\Lambda$  & $Ra_c(\times 10^5)$  & $k_x^{c}$  & $Ra_c(\times 10^5)$  & $k_x^{c}$ & $Ra_c(\times 10^5)$  & $k_x^{c}$ \\
\hline
& 0	   &15.14221  &25 &13.12945  &26 &8.12454 &27\\
&0.001 &15.07739  &25 &13.12914  &26 &8.12396 &27\\
&0.01  &15.07218  &25 &13.10094  &25 &8.11457 &26\\
&0.1   &5.71154   &25 &12.79451  &25 &7.94515 &26\\ 
&1     &5.53475   &4  &5.19500   &4  &3.74895 &5\\
&2	   &4.44025   &3  &4.22300   &4  &3.23055 &4\\
&3	   &3.26015   &3  &3.16050   &4  &2.56552 &4\\
&4     &2.69498   &4  &2.65600   &4  &2.24651 &4\\
&5	   &2.37949   &4  &2.40100   &4  &2.08451 &4\\
&10	   &2.06552   &4  &2.49250   &4  &2.24542 &4\\
\hline
\multirow{11}{*}{\begin{turn}{90}$E = 10^{-5}$\end{turn}} & $\Lambda$  & $Ra_c(\times 10^6)$  & $k_x^{c}$  & $Ra_c(\times 10^6)$  & $k_x^{c}$ & $Ra_c(\times 10^6)$  & $k_x^{c}$ \\
\hline
& 0	   &34.487322 &55 &29.500124  &57 &18.047152  &60\\
&0.001 &34.411297 &55 &29.499845  &57 &18.035247  &60\\
&0.01  &34.352425 &55 &29.474641  &57 &18.076451  &60\\
&0.1   &34.011498 &5  &29.544153  &56 &17.654194  &59\\ 
&1     &5.700019  &4  &5.334001   &4  &3.886991   &5\\
&2	   &4.438175  &3  &4.269899   &3  &3.282515   &4\\
&3	   &3.092505  &3  &3.035004   &3  &2.499998   &4\\
&4     &2.451205  &3  &2.439009   &3  &2.101755   &4\\
&5	   &2.090505  &3  &2.124998   &3  &1.880501   &4\\
&10	   &1.579152  &3  &2.040101   &3  &1.868775   &4\\
\hline
\label{Table_q10}
\end{tabular}
\end{table}

In previous investigations, on thermal convection\cite{garai2022convective, barman2024role} and magnetoconvection with non-uniformly imposed magnetic field\cite{sreenivasan2024oscillatory}, in presence of stable stratification showed that $Ra_c^h$ is typically less than the $Ra_c^{\infty}$. However, at high Roberts number ($q >> 1$) and strongly imposed magnetic field regime, the critical Rayleigh number with stable stratification ($Ra_c^h$) may become higher than the reference case of fully unstable stratification\cite{sreenivasan2024oscillatory}. This highlights that the effect of stable stratification on convective onset is strongly sensitive to magnetic field strength and diffusivity ratio. To investigate this, we consider $E = 10^{-4}$, $\Lambda = 0, 0.01$ for weakly imposed magnetic field regime, and $\Lambda = 1, 5$ for strongly imposed magnetic field regime, with a fixed diffusivity ratio $q = 1$. The corresponding critical Rayleigh number ($Ra_c$) are shown in Table (\ref{Table_q1}). The influence of slow to rapid rotation, and low to high diffusivity ratios on critical values of convective onset are explored in greater detail in preceding section (\ref{sec3a3}).  

For the non-magnetic case ($\Lambda = 0$) and weakly imposed magnetic field ($\Lambda = 0.01$), the relative change in critical Rayleigh number, $D_{Ra}(\%)$, is approximately $\sim14\%$ for weak stratification ($h = 0.8$) and increases to around $\sim 47\%$ for strong stratification ($h = 0.6$). This increase is attributed to the enhanced heat flux accommodated in the unstable region ($z < h$), where $|\partial T / \partial z| = 1.8$ for $h = 0.6$, compared to $1.0672$ for $h = 0.8$ (see Table \ref{Table_models}). In the magnetically dominated regime ($\Lambda \geq 1$), $D_{Ra}$ reduces to $\sim 6\%$ and $\sim 32\%$ for $h = 0.8$ and $h = 0.6$, respectively, under a moderately strong axial magnetic field ($\Lambda = 1$), and slightly increases to $\sim 7\%$ and $\sim35\%$ under stronger magnetic forcing ($\Lambda = 5$). It indicates that stable stratification leads to a drop in the critical Rayleigh number, with a more pronounced reduction in non-magnetic and weak-field cases, while the effect becomes less significant under strong magnetic fields due to magnetic suppression of convection when $q = 1$. 

The critical horizontal wave number ($k_c^x$) characterizing the convective length scale also alters in presence of stable stratification. For fully unstable stratification ($h = \infty$, Fig. \ref{f_SSL_contour}a) and weakly stratified case ($h = 0.8$, Fig. \ref{f_SSL_contour}i), the horizontal wavenumber is $k_c^x = 25$ at $\Lambda = 0$, increasing to $k_c^x = 27$ under strong stratification ($h = 0.6$, Fig. \ref{f_SSL_contour}q). With a weak magnetic field ($\Lambda = 0.01$), the wavenumber remains unchanged from the non-magnetic case across all stratification levels: $k_c^x = 25$ for $h = \infty$ and $h = 0.8$ [Figs. \ref{f_SSL_contour}b, j], and increases to $k_c^x = 27$ for $h = 0.6$ (Fig. \ref{f_SSL_contour}r), indicating a slight magnetic modification of convective length scale. At $\Lambda = 1$, the horizontal wavenumber ($k_c^x$) is 4 for both the fully unstable stratification ($h = \infty$, Fig. \ref{f_SSL_contour}c) and weakly stably stratified ($h = 0.8$, Fig. \ref{f_SSL_contour}k) cases, but increases to $k_c^x = 5$ for strongly stable stratification ($h = 0.6$, Fig. \ref{f_SSL_contour}s). This indicates that strong stratification can influence convective length scales at $\Lambda = 1$, whereas a weak stable layer does not. Moreover, the lower wavenumber compared to the rotation-dominated regime signifies larger, thicker convection rolls at this intermediate magnetic field strength. With a further increase to $\Lambda = 5$, the wavenumber rises across all cases—$k_c^x = 7$ for $h = \infty$, 8 for $h = 0.8$, and 9 for $h = 0.6$—showing that both weak and strong stratification significantly affect the convective length scale under strong magnetic fields. 


\subsubsection{Variation of $Ra_c$ and $k_c^x$ with $\Lambda$} \label{sec3a3}
\begin{figure}[htbp]
    \centering
    \includegraphics[clip, trim=0cm 6cm 0cm 0cm, width=1\textwidth]{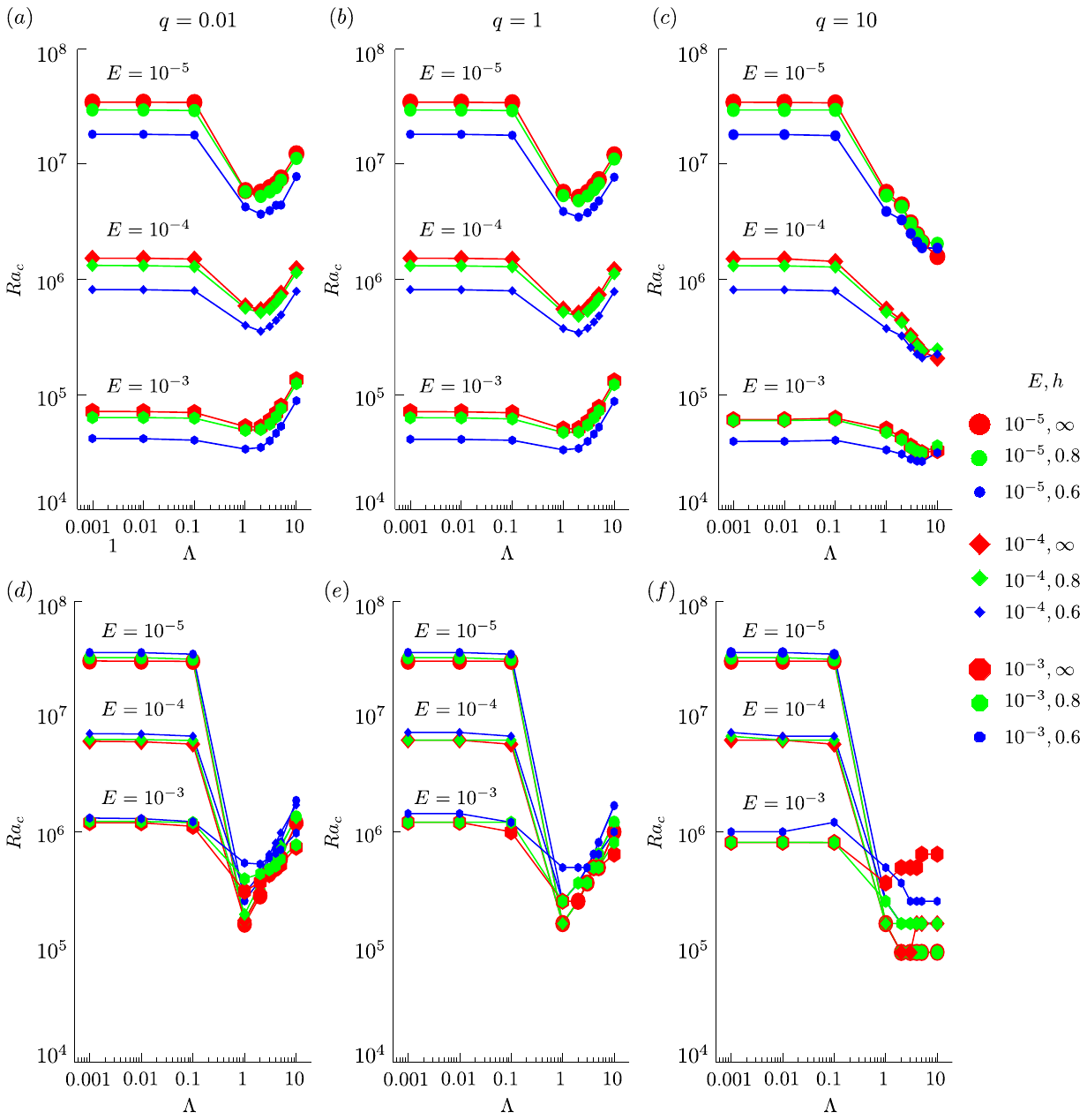}
    \caption{Plots of the critical Rayleigh number ($Ra_c$) and horizontal wavenumber ($k_c^x$) as functions of $\Lambda$. Panels (a), (b), and (c) show $Ra_c$ vs. $\Lambda$ for $q = 0.01$, $q = 1$, and $q = 10$, respectively. Similarly, panels (d), (e), and (f) show $k_c^x$ vs. $\Lambda$ for the same values of $q$. Each marker represents a different rotational regime: filled circles for rapid rotation ($E = 10^{-5}$), filled diamonds for moderate rotation ($E = 10^{-4}$), and filled hexagons for slow rotation ($E = 10^{-3}$). Colors denote stratification strength: red for fully unstable stratification ($h = \infty$), green for weak stable stratification ($h = 0.8$), and blue for strong stable stratification ($h = 0.6$).}
    \label{f_profile_q_variation}
\end{figure}

In the previous section, we analyzed how partial stable stratification affects the critical Rayleigh number ($Ra_c$) and horizontal wavenumber ($k_c^x$) at fixed rotation ($E = 10^{-4}$) and diffusivity ratio ($q = 1$). The results show that stable stratification lowers $Ra_c$, thereby promoting earlier convection onset, while simultaneously increasing $k_c^x$, which favors the formation of smaller-scale structures. Extending this analysis, we now explore variations in magnetic field strength ($\Lambda$, from 0 to 10) and diffusivity ratio ($q$, from 0.01 to 10), with results presented in Fig.(\ref{f_profile_q_variation}) for different rotation rates ($E = 10^{-5}, 10^{-4}, 10^{-3}$). To quantitatively interpret the variation of $Ra_c$ and $k_c^x$ with magnetic field strength ($\Lambda$), scaling laws of the form
\begin{equation}
    Ra_c \propto \alpha_1 \Lambda^{\beta_1},
\end{equation}
and
\begin{equation}
    k_c^x \propto \alpha_2 \Lambda^{\beta_2}, 
\end{equation} 
are derived using linear regression on data from Tables (\ref{Table_q0.01}) – (\ref{Table_q10}) for $q = 0.01$, 1, and 10. Negative exponents ($\beta_1, \beta_2 < 0$) indicate inverse scaling with $\Lambda$, and high (low) values of pre-factors ($\alpha_1, \alpha_2$) indicates high (low) values of $Ra_c$ and $k_c^x$.  

In the weak field regime ($\Lambda < 1$), $Ra_c$ varies slowly with increasing $\Lambda$ (0.001–0.1) across all $E$ and $q$ (Figs.\ref{f_profile_q_variation}a–c; Tables\ref{Table_q0.01}–\ref{Table_q10}), but decreases notably with slower rotation (higher $E$), due to reduced rotational constraints—from $E = 10^{-5}$ (filled circles) to $E = 10^{-3}$ (filled hexagons). For $q = 0.01$, $1$, and $10$, $Ra_c$ is consistently lower for weak ($h = 0.8$) and strong ($h = 0.6$) stratification than for the fully unstable case ($h = \infty$), driven by enhanced superadiabatic gradients in the unstable region ($0 < z < h$) [Table\ref{Table_models}]. Meanwhile, $k_c^x$ is largely unaffected by $\Lambda$ but increases with stratification strength (Figs.\ref{f_profile_q_variation}d–f), indicating smaller-scale convection. Scaling laws confirm a weak inverse dependence on $\Lambda$ ($Ra_c, k_c^x \propto \Lambda^{-0.01}$) across all $E$, $h$, and $q = 0.01, 1$. For rapid rotation ($E = 10^{-5}$), the pre-factors for $Ra_c$ and $k_c^x$ slightly decrease and increase, respectively, with stratification—from $Ra_c \propto 7.51\Lambda^{-0.01}$ to $\propto 7.24\Lambda^{-0.01}$ and $k_c^x \propto 1.72\Lambda^{-0.01}$ to $1.75$ (for $h = \infty$ to $h = 0.6$). With the similar trends at $E = 10^{-3}$ the pre-factor reduces indicating weakening of rotation. For high diffusivity ratio ($q = 10$), the scaling exponent for $Ra_c$ remains unchanged at $-0.01$, consistent with lower $q$ values ($\leq 1$), across all $E$ and $h$, though the pre-factors decrease slightly (by ~0.5\%). The $Ra_c$ and $k_c^x$ scalings at $E = 10^{-5}$ and $10^{-4}$ are nearly identical to those for $q = 0.01, 1$, but under slow rotation ($E = 10^{-3}$), the $k_c^x$ pre-factor reduces to $0.95$ for $h = \infty$ and $0.8$, and further to $0.97$ for $h = 0.6$, as shown in Fig.\ref{f_profile_q_variation}f (blue hexagons).Overall, in the weak field regime, stable stratification primarily governs the reduction in $Ra_c$ and the rise in $k_c^x$, while the effects of $\Lambda$, $E$, and $q$ are comparatively minor—aligning with previous results for rotating magnetoconvection with fully unstable stratification \cite{aujogue2015onset}.   

Transitioning from the weak field regime ($\Lambda < 1$) to the strong field regime ($\Lambda \geq 1$) causes a sudden drop in both the critical Rayleigh number ($Ra_c$) and critical horizontal wavenumber ($k_c^x$), as shown in Figs. (\ref{f_profile_q_variation}a–c) and (\ref{f_profile_q_variation}d–f), respectively. This drop results from the relaxation of rotational constraints as the influence of the imposed magnetic field becomes significant. The sharpness of this transition increases with stronger rotation (i.e., lower $E$), becoming more pronounced as $E$ decreases from $10^{-3}$ to $10^{-5}$, regardless of diffusivity ratio ($q$) or stratification ($h$). However, for low and moderate diffusivity ratios ($q = 0.01, 1$), both $Ra_c$ and $k_c^x$ begin to increase again as $\Lambda$ exceeds unity, indicating a regime shift near $\Lambda = 1$ from rotationally dominated (viscous) to magnetically dominated dynamics—a behavior consistent with the mode-switching scenario described by Aujogue et al.\cite{aujogue2015onset}. Under rapid rotation ($E = 10^{-5}$) and fully unstable stratification ($h = \infty$), scaling shows $Ra_c \propto 6.66 \Lambda^{0.32}$ and $k_c^x \propto 0.59 \Lambda^{0.44}$, which modify to $Ra_c \propto 6.46 \Lambda^{0.26}$ and $k_c^x \propto 0.65 \Lambda^{0.44}$ under strong stable stratification ($h = 0.6$). Under slower rotation ($E = 10^{-3}$), the scalings become $Ra_c \propto 4.62 \Lambda^{0.40}$ and $k_c^x \propto 0.70 \Lambda^{0.20}$ for unstable stratification, and $Ra_c \propto 4.44 \Lambda^{0.41}$ and $k_c^x \propto 0.81 \Lambda^{0.13}$ for strong stable stratification. The positive exponents in all these cases indicate that both $Ra_c$ and $k_c^x$ rise with increasing magnetic field strength in the strong field regime ($\Lambda = 1 \rightarrow 10$), as evident in Figs. (\ref{f_profile_q_variation}a–b) and (\ref{f_profile_q_variation}d–e). This increase in $Ra_c$ is attributed to the stabilizing influence of the Lorentz force, while the increase in $k_c^x$ reflects the restoration of columnar flow structures due to the elongation of axial velocity along the direction of the imposed magnetic field. Overall, in the weak field regime ($\Lambda < 1$), $Ra_c$ remains nearly constant and is primarily reduced by stratification and weakened rotation, while $k_c^x$ increases modestly with stratification, showing minimal sensitivity to $\Lambda$, $E$, and $q$. As $\Lambda$ exceeds unity, a sharp drop in $Ra_c$ and $k_c^x$ marks a transition to the strong field regime, followed by a gradual increase due to the stabilizing Lorentz force and restoration of columnar convection, particularly under rapid rotation.

For high diffusivity ratio ($q = 10$), the scaling behavior of $Ra_c$ and $k_c^x$ in the weak field regime ($\Lambda < 1$) remains nearly identical to that observed for $q = 0.01$ and $1$. However, this trend changes markedly in the strong field regime, as evident from Figs. (\ref{f_profile_q_variation}c) and (\ref{f_profile_q_variation}f). Here, the critical Rayleigh number ($Ra_c$) decreases monotonically with increasing $\Lambda$ (from 1 to 10), regardless of rotation rate, and a similar decreasing trend is observed for $k_c^x$, highlighting the distinct behavior of high-$q$ systems under strong magnetic fields. Under rapid rotation ($E = 10^{-5}$), the scalings are $Ra_c \propto 6.71 \Lambda^{-0.56}$ and $k_c^x \propto 0.56 \Lambda^{-0.12}$ for fully unstable stratification, which modify to $Ra_c \propto 6.50 \Lambda^{-0.32}$ and $k_c^x \propto 0.67 \Lambda^{-0.32}$ for strong stable stratification ($h = 0.6$). Under slow rotation ($E = 10^{-3}$), the scalings become $Ra_c \propto 4.62 \Lambda^{-0.43}$ and $k_c^x \propto 0.81 \Lambda^{0.12}$ for unstable stratification, and $Ra_c \propto 4.43 \Lambda^{-0.02}$ and $k_c^x \propto 0.80 \Lambda^{-0.15}$ for strong stratification. The negative exponents for both $Ra_c$ and $k_c^x$ contrast with the positive exponents seen for $q = 0.01$ and $1$, indicating that, in high-$q$ regimes, convection onsets at lower thresholds as thermal diffusion dominates magnetic diffusion. In this case, the Lorentz force enhances, rather than suppresses, convection. Additionally, the decrease in $k_c^x$ with increasing $\Lambda$ suggests suppression of small-scale features in the strong field regime at high diffusivity ratios.  

In the weak field regime ($\Lambda < 1$), the critical Rayleigh number ($Ra_c$) remains nearly constant across all $q$ and $E$, with stratification and weaker rotation reducing $Ra_c$ and increasing $k_c^x$. A sharp drop in both $Ra_c$ and $k_c^x$ occurs at $\Lambda \approx 1$, marking a transition to the strong field regime. For $q = 0.01$ and $1$, $Ra_c$ and $k_c^x$ increase with $\Lambda$ in this regime due to the stabilizing Lorentz force and restoration of columnar convection. In contrast, for high diffusivity ($q = 10$), both $Ra_c$ and $k_c^x$ decrease with increasing $\Lambda$, indicating enhanced convection and suppression of small-scale features as thermal diffusion dominates magnetic effects.



\subsection{Impact of rotation ($E$) and diffusivity ratio ($q$)}\label{sec3b}
    
       

The interplay of background rotation, an imposed axial magnetic field, and stable stratification leads to complex convective dynamics even at the onset of convection. To assess the influence of rotation, the Ekman number is varied from $10^{-3}$ to $10^{-5}$, with the Roberts number fixed at $q = 1$. Both fully unstable ($h = \infty$) and strong stable ($h = 0.6$) stratification cases are examined to investigate the combined effects of rotation and stratification under weak ($\Lambda = 0.01$) and strong ($\Lambda = 5$) magnetic field regimes. 

\subsubsection{Variation with rotation rate ($E$)}\label{sec3b1}
\begin{figure}[htbp]
    \centering
    \includegraphics[clip, trim=2.5cm 15.5cm 3cm 0cm, width=1\textwidth]{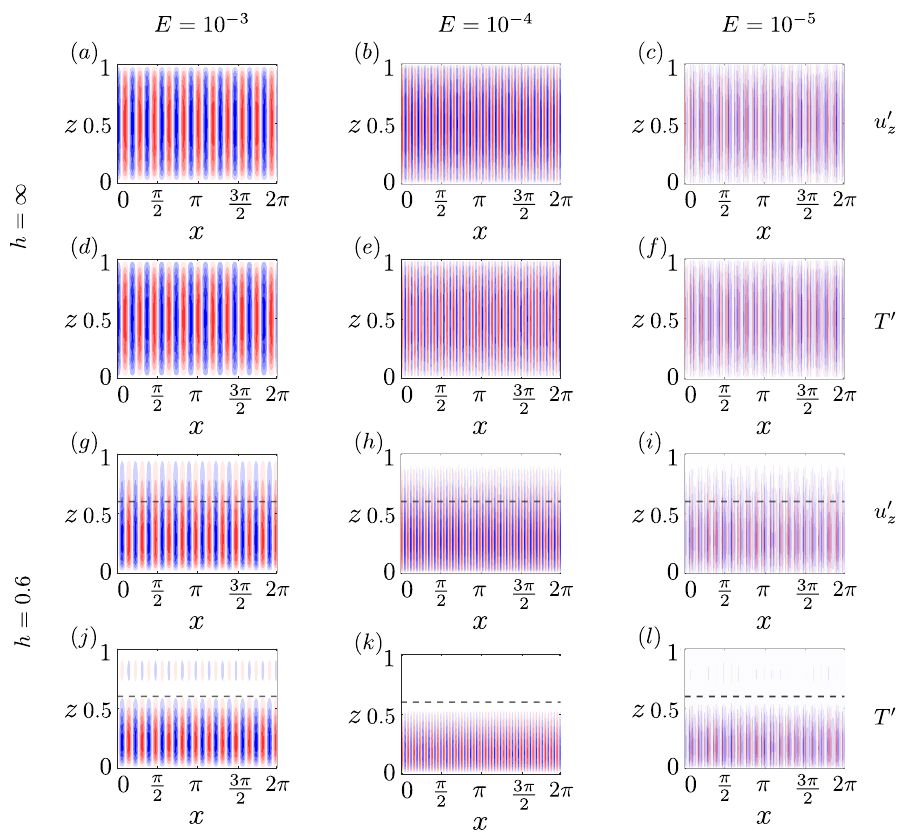}
    \caption{Contour plots of axial velocity perturbation ($u_z^{\prime}$) and temperature perturbation ($T^{\prime}$) for rotationally dominated regime ($\Lambda = 0.01$) at $q = 1$. Contours of $u_z^{\prime}$ : (a)-(c)  for $h = \infty$ and (g)-(i) for $h = 0.6$ (dashed line at $z = 0.6$). Contours of $T^{\prime}$ : (d)-(f) for $h = \infty$ and (j)-(l) for $h = 0.6$ (dashed line at $z = 0.6$). The first, second, and third columns indicate the slow rotation ($E = 10^{-3}$), moderate rotation ($E = 10^{-4}$), and rapid rotation cases ($E = 10^{-5}$), respectively.}
    \label{f_lowEs_contour_rotation}
\end{figure}

In the weak magnetic field regime ($\Lambda = 0.01$), elongated columnar rolls emerge. For fully unstable stratification ($h = \infty$), temperature perturbations closely follow the axial velocity contours, as seen by comparing figures (\ref{f_lowEs_contour_rotation}d–f) with (\ref{f_lowEs_contour_rotation}a–c). As the rotation rate increases (i.e., $E$ decreases from $10^{-3}$ to $10^{-5}$), the columns become thinner due to the increasing dominance of rotational constraint under geostrophic balance (figures \ref{f_lowEs_contour_rotation}a–c). This enhanced columnarity is reflected in higher horizontal wavenumbers, indicating smaller-scale features. For $h = \infty$, the wavenumber at moderate rotation is approximately $25/11$ times that at slow rotation, and about $55/11$ times greater at rapid rotation ($E = 10^{-5}$). Convection onset is delayed with increasing rotation, as shown by the rise in $Ra_c$ in Table (\ref{Table_q1}), due to enhanced system stability and stronger Taylor–Proudman constraint \cite{barman2024role, garai2022convective}. Under strong stable stratification ($h = 0.6$), axial symmetry breaks down, resulting in asymmetric convective modes (figures \ref{f_lowEs_contour_rotation}g–l). Temperature perturbations remain confined within the unstable region (below the dashed line in figures \ref{f_lowEs_contour_rotation}j–l), while axial velocity penetrates the stable layer (figures \ref{f_lowEs_contour_rotation}g–i), a hallmark of rotating penetrative convection \cite{garai2022convective, barman2025penetration}. In addition for strong stable stratification ($h = 0.6$), the horizontal wavenumber ($k_c^x$) increases by approximately $27/12$ at moderate rotation and $60/12$ at rapid rotation, compared to slow rotation. Relative to the $h = \infty$ case, $k_c^x$ increases slightly (by $\sim8\%$) for all rotation rates under strong stratification, as shown in Table (\ref{Table_q1}). Alongside reduced convective length scales, $Ra_c$ rises with rotation in both $h = \infty$ and $h = 0.6$ cases: increasing by $\sim95\%$ from $E = 10^{-3}$ to $10^{-4}$ and by $\sim99\%$ from $E = 10^{-3}$ to $10^{-5}$. 

\begin{figure}[htbp]
    \centering
    \includegraphics[clip, trim=2.5cm 15.5cm 3cm 0cm, width=1\textwidth]{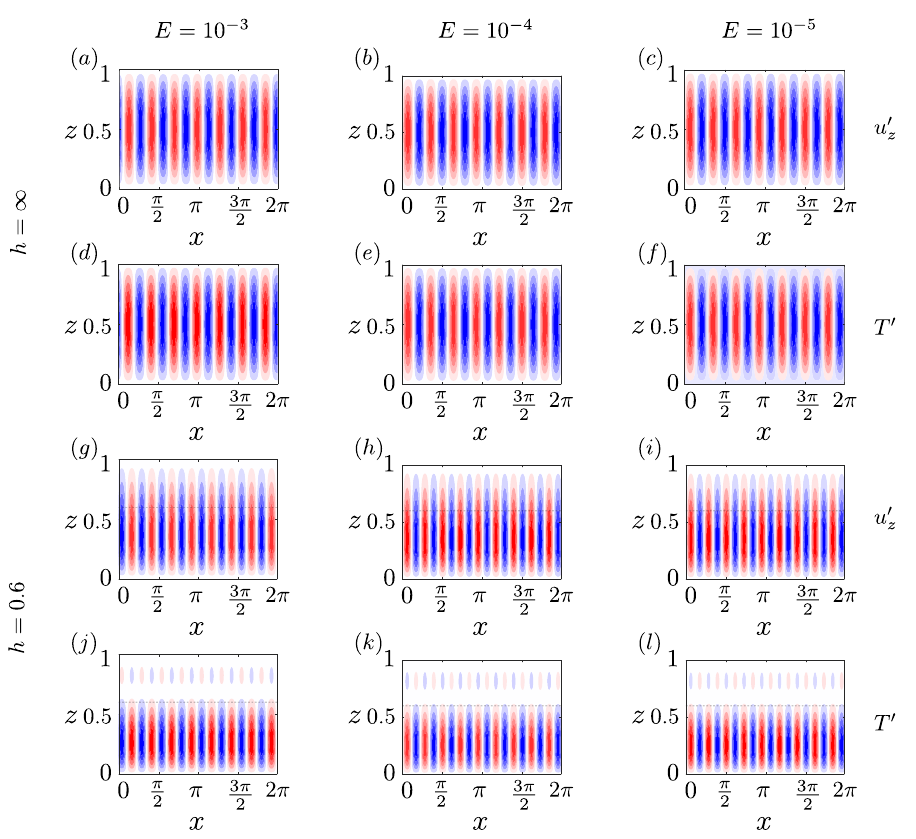}
    \caption{Contour plots of axial velocity perturbation ($u_z^{\prime}$) and temperature perturbation ($T^{\prime}$) for magnetically dominated regime ($\Lambda = 5$) at $q = 1$. Contours of $u_z^{\prime}$ : (a)-(c)  for $h = \infty$ and (g)-(i) for $h = 0.6$ (dashed line at $z = 0.6$). Contours of $T^{\prime}$ : (d)-(f) for $h = \infty$ and (j)-(l) for $h = 0.6$ (dashed line at $z = 0.6$). The first, second, and third columns indicate the slow rotation ($E = 10^{-3}$), moderate rotation ($E = 10^{-4}$), and rapid rotation cases ($E = 10^{-5}$), respectively.}
    \label{f_highEs_contour_rotation}
\end{figure}

In contrast to the weak field regime ($\Lambda = 0.01$), the strong field regime ($\Lambda = 5$) disrupts columnarity induced by rotation, resulting in thicker convection rolls, as seen in Figs. \ref{f_highEs_contour_rotation}a–c for fully unstable stratification ($h = \infty$). Nonetheless, temperature perturbations still align closely with axial velocity contours (Figs. \ref{f_highEs_contour_rotation}d–f). Unlike the weak field case, the critical horizontal wavenumber ($k_c^x$) remains constant at 7 across all rotation rates ($E = 10^{-3}$ to $10^{-5}$), indicating minimal rotational influence. Note that the columnar rolls observed in the magnetically dominated regime ($\Lambda > 1$) arise from flow elongation along the direction of the imposed axial magnetic field via magnetic stretching, rather than from rotational effects. Under strong stable stratification ($h = 0.6$), axial symmetry again breaks down, as seen in figures (\ref{f_highEs_contour_rotation}g–j). Axial velocity ($u_z^{\prime}$) and temperature perturbations ($T^{\prime}$) become decoupled, as $T^{\prime}$ is suppressed while $u_z^{\prime}$ penetrates into the stable stratification [Figs. \ref{f_highEs_contour_rotation}g–i]. Interestingly, short rolls in opposite phase to those in the unstable layer appear within the stable stratified region ($0.6 < z < 1$) in the temperature perturbations [Figs. \ref{f_highEs_contour_rotation}j–l], regardless of rotation rate. Their presence, also observed under slow rotation in the weak field regime [Fig. \ref{f_highEs_contour_rotation}j], may indicate wave-like phenomena, consistent with earlier studies on penetrative thermal convection \cite{garai2022convective, barman2024role}. Moreover, the critical wave number ($k_c^x$) increases from 7 (for $h = \infty$) to 8 under slow rotation ($E = 10^{-3}$), and further to 9 for higher rotation rates ($E = 10^{-4}, 10^{-5}$). This contrasts with the fully unstable case, where $k_c^x$ remains unchanged. Despite weak dependence of $k_c^x$ on rotation in the strong field regime, the critical Rayleigh number ($Ra_c$) still increases significantly—by approximately $\sim90\%$ from $E = 10^{-3}$ to $10^{-4}$ and by $\sim99\%$ from $E = 10^{-3}$ to $10^{-5}$—for both stratification models.

      
Rotation delays the onset of convection across both weak and strong magnetic field regimes, and for both fully unstable and strongly stratified models. To assess the influence of stable stratification at each rotation rate, the relative percentage drop in critical Rayleigh number ($D_{Ra}\%$) is computed using equation (\ref{eq_relative_drop_Ra}). In the weak field regime ($\Lambda = 0.01$), $D_{Ra}\%$ is approximately $\sim 42\%$ for slow rotation ($E = 10^{-3}$), increasing slightly to $\sim 45\%$ for moderate and rapid rotation. In contrast, for the strong field regime ($\Lambda = 5$), $D_{Ra}\%$ is $\sim32\%$ for slow rotation and rises marginally to $\sim 34 \%$ at higher rotation rates. This indicates that the stabilizing effect of stratification is more pronounced in the weak field case, whereas convection sets in earlier under strong magnetic fields.

Furthermore, a close inspection into the contours of convective instabilities, for strong stable layer ($h = 0.6$), provides a qualitative understanding how the suppression (of $T^{\prime}$) and penetration (by $u_z^{\prime}$) evolve with rotation in weak and strong field regimes. The axial velocity contours [Figs. \ref{f_lowEs_contour_rotation} for $h = 0.6$] reveals that the penetration above the stable layer decreases—as indicated by the shrinking gap between the top plate and the peak of convection rolls—as the rotation rate increases from slow to rapid [compare Figs. \ref{f_lowEs_contour_rotation}g, \ref{f_lowEs_contour_rotation}h, \ref{f_lowEs_contour_rotation}i]. Similarly, the suppression of temperature perturbations strengthens with increasing rotation, as indicated by the widening gap between the interface height (dashed line) and the peak of temperature perturbation within the unstable layer, evident from the comparison of slow rotation (Fig. \ref{f_lowEs_contour_rotation}j) with moderate and rapid rotation (Figs. \ref{f_lowEs_contour_rotation}k and \ref{f_lowEs_contour_rotation}l). It qualitatively indicates that the penetration of convective instabilities reduces as rotation rate increases in weak field regime. However, under strong magnetic field regime ($\Lambda = 5$), the suppression and penetration effects becomes almost invariant with rotation as depicted in figure (\ref{f_highEs_contour_rotation}g - i) for penetration by $u_z^{\prime}$ and (\ref{f_highEs_contour_rotation}j - l) for suppression of $T^{\prime}$. Therefore, it is inferred qualitatively that rotation enhances stratification-induced stabilization more in weak than strong magnetic fields, with convective penetration and temperature suppression becoming increasingly rotation-dependent only in the weak field regime. For an enhanced understanding, the detailed quantitative estimates of penetration depth is discussed in section (\ref{sec3c}).



\subsubsection{Variation with diffusivity ratio ($q$)}\label{sec3b2}
Understanding the interaction among stratification, rotation, and magnetic field is essential for characterizing the onset of convection. Inherently, the magnetohydrodynamic system involves three diffusion mechanisms: viscous diffusivity ($\nu$) for momentum, thermal diffusivity ($\kappa$) for heat, and magnetic diffusivity ($\eta$) for magnetic field dissipation. To explore the impact of their interplay through the diffusivity ratio ($q$), we explore particular parameter regimes obtained by fixing the rotation rate at $E = 10^{-4}$ and considering two magnetic field strengths: weak ($\Lambda = 0.01$) and strong ($\Lambda = 5$). In each of the above regimes, two stratification profiles are examined: fully unstable ($h = \infty$) and strongly stable ($h = 0.6$).

\begin{figure}[htbp]
    \centering
    \includegraphics[clip, trim=2.5cm 15cm 3cm 0cm, width=1\textwidth]{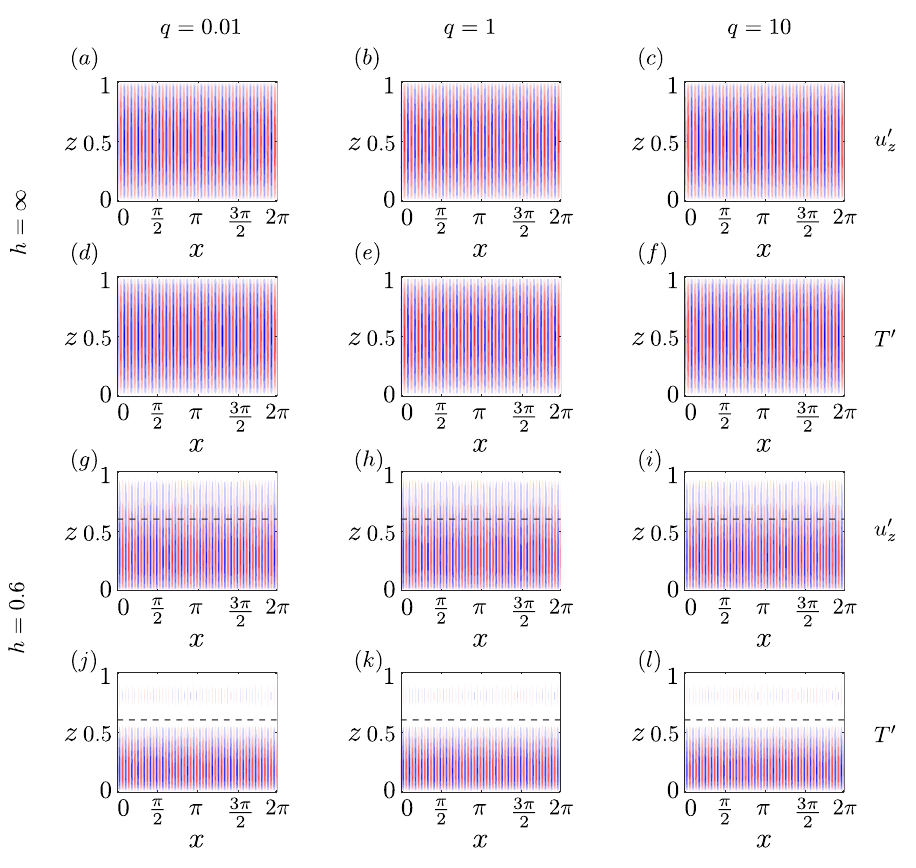}
    \caption{Contour plots of axial velocity perturbation ($u_z^{\prime}$) and temperature perturbation ($T^{\prime}$) for rotationally dominated regime ($\Lambda = 0.01$) at $E = 10^{-4}$. Contours of $u_z^{\prime}$ : (a)-(c)  for $h = \infty$ and (g)-(i) for $h = 0.6$ (dashed line at $z = 0.6$). Contours of $T^{\prime}$ : (d)-(f) for $h = \infty$ and (j)-(l) for $h = 0.6$ (dashed line at $z = 0.6$). The first, second, and third columns indicate the low diffusivity ratio ($q = 0.01$), moderate diffusivity ratio ($q = 1$), and high diffusivity ratio ($q = 10$), respectively.}
    \label{f_lowEs_contour_q}
\end{figure}

In the weak field regime ($\Lambda = 0.01$) with fully unstable stratification ($h = \infty$), elongated, axially symmetric columnar rolls appear consistently across low to high Roberts numbers ($q$) [Figs. \ref{f_lowEs_contour_q}a–c], with temperature perturbations ($T^{\prime}$) closely following the axial velocity structures ($u_z^{\prime}$) [Figs. \ref{f_lowEs_contour_q}d–f]. Introducing strong stable stratification ($h = 0.6$) breaks the mid-plane symmetry, confines $T^{\prime}$ to the unstable layer [Figs. \ref{f_lowEs_contour_q}g–i], and allows partial penetration of $u_z^{\prime}$ into the stable region [Figs. \ref{f_lowEs_contour_q}j–l]. These observations suggest that variation in $q$ has minimal effect on the spatial structure of convective instabilities. This is supported by the nearly unchanged critical horizontal wavenumber, which remains at $k_c^x = 25$ for $h = \infty$ and increases slightly to 26 for $h = 0.6$—a modest $\sim 4\%$ rise—across $q = 0.01$, 1, and 10 (see Tables \ref{Table_q0.01}, \ref{Table_q1}, and \ref{Table_q10} for $E = 10^{-4}$, $\Lambda = 0.01$). Similarly, the critical Rayleigh number ($Ra_c$) shows negligible variation with $q$: decreasing by only $\sim 0.1\%$ from $q = 0.01$ to 1, and by $\sim 1.2\%$ from $q = 1$ to 10, for both stratification cases. In addition, the relative reduction in $Ra_c$ due to stable stratification ($D_{Ra}\%$) remains nearly constant at $\sim 47\%$ across all $q$ values, indicating that variation in Roberts number ($q$) has minimal impact on the spatial structure and onset threshold of convection.  

\begin{figure}[htbp]
    \centering
    \includegraphics[clip, trim=2.5cm 15cm 3cm 0cm, width=1\textwidth]{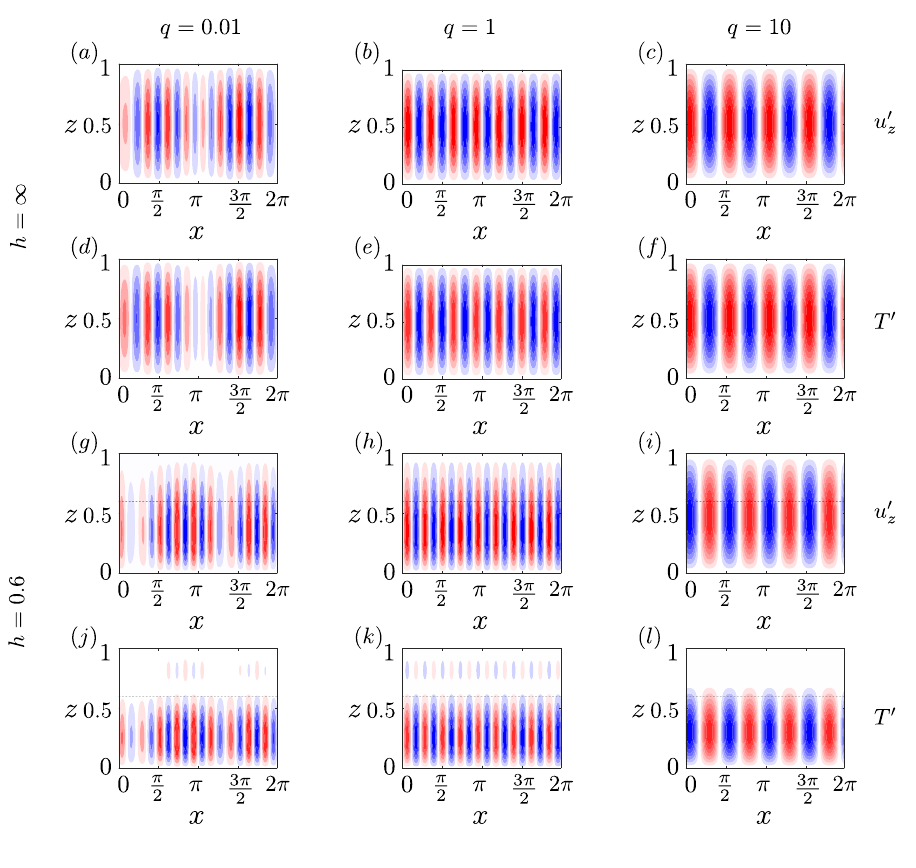}
    \caption{Contour plots of axial velocity perturbation ($u_z^{\prime}$) and temperature perturbation ($T^{\prime}$) for magnetically dominated regime ($\Lambda = 5$) at $E = 10^{-4}$. Contours of $u_z^{\prime}$ : (a)-(c)  for $h = \infty$ and (g)-(i) for $h = 0.6$ (dashed line at $z = 0.6$). Contours of $T^{\prime}$ : (d)-(f) for $h = \infty$ and (j)-(l) for $h = 0.6$ (dashed line at $z = 0.6$). The first, second, and third columns indicate the low diffusivity ratio ($q = 0.01$), moderate diffusivity ratio ($q = 1$), and high diffusivity ratio ($q = 10$), respectively.}
    \label{f_highEs_contour_q}
\end{figure}

In the strong field regime ($\Lambda = 5$) with fully unstable stratification ($h = \infty$), convection rolls appear thicker [Figs. \ref{f_highEs_contour_q}a–c] compared to the weak field case ($\Lambda = 0.01$), and their structure varies with diffusivity ratio ($q$). At higher $q( = 10)$ values , the rolls are notably thicker [Figs. \ref{f_highEs_contour_q}c, f] than at lower $q$  as shown in figures (\ref{f_highEs_contour_q}a, d) for $q = 0.01$ and (\ref{f_highEs_contour_q} b, e) for $q = 1$. When a strong stable stratification ($h = 0.6$) is imposed, the convective rolls gradually thicken with increasing $q$, become axially asymmetric, and partially penetrate into the stable layer [Figs. \ref{f_highEs_contour_q}g–i]. Temperature perturbations remain confined within the unstable layer for $q = 0.01$ and $1$ [Figs. \ref{f_highEs_contour_q}j, k], while a slight penetration occurs at $q = 10$ [Fig. \ref{f_highEs_contour_q}l]. This thickening trend with increasing $q$ is reflected in the critical horizontal wave number ($k_c^x$), which decreases from 8 at $q = 0.01$ to 7 at $q = 1$, and to 4 at $q = 10$ for $h = \infty$ (see tables \ref{Table_q0.01}–\ref{Table_q10} for $E = 10^{-4}$, $\Lambda = 5$). Stable stratification ($h = 0.6$) increases $k_c^x$ compared to the unstable case but retains the same descending trend with $q$: $k_c^x = 10$ for $q = 0.01$, 9 for $q = 1$, and 4 for $q = 10$. 

Similarly, the critical Rayleigh number ($Ra_c$) decreases with increasing $q$ for both stratification models. From $q = 0.01$ to $q = 1$, $Ra_c$ drops by $\sim 3\%$, and from $q = 0.01$ to $q = 10$, the drop becomes $\sim 69\%$ for $h = \infty$ and $\sim 58\%$ for $h = 0.6$, which is significantly more than in the weak field regime. The relative drop in $Ra_c$ due to stable stratification ($D_{Ra}(\%)$) remains $\sim35 \%$ for low to moderate $q$ (= 0.01 – 1) values, but reduces sharply to $\sim 12\%$ for $q = 10$ in the magnetically dominated regime. This indicates that the weak magnetic diffusion diminishes the effect of stable stratification in the magnetically dominated regime under strong magnetic field. Therefore, it has been inferred that, in the weak field regime ($\Lambda = 0.01$), variation in diffusivity ratio ($q$) has negligible impact on convective structure and onset thresholds, while in the strong field regime ($\Lambda = 5$), increasing $q$ significantly thickens convection rolls and lowers both $k_c^x$ and $Ra_c$, reducing the stabilizing influence of stratification at high $q$.

Furthermore, a close examination of convective instability contours in the presence of a strong stable layer offers qualitative insights into how penetration (by $u_z^{\prime}$) and suppression (of $T^{\prime}$) evolve with diffusivity ratio across weak and strong magnetic field regimes. In the weak field case ($\Lambda = 0.01$), axial velocity contours indicate that the extent of penetration above the interface height ($h$) remains nearly unchanged as the diffusivity ratio ($q$) increases [Figs.\ref{f_lowEs_contour_q}g–i]. Similarly, temperature suppression—evidenced by the consistent gap between the interface height (dashed line) and the location of peak temperature perturbation—remains invariant across $q$ values [Figs.\ref{f_lowEs_contour_q}j–l]. This trend persists in the strong field regime ($\Lambda = 5$), where both penetration and suppression remain largely unaffected by changes in $q$, as seen in the axial velocity [Figs.\ref{f_highEs_contour_q}g–i] and temperature perturbation [Figs.\ref{f_highEs_contour_q}j–l] contours. Thus, it indicates that penetration of axial flow and suppression of temperature perturbations across the stable layer remain largely insensitive to change in diffusivity ratio ($q$) in both weak and strong magnetic field regimes. 

\subsection{Depth of penetration}\label{sec3c}
In Section \ref{sec3a1}, we qualitatively observed that the penetration depth of axial velocity ($u_z^{\prime}$) into the stable layer varies with magnetic field strength ($\Lambda$): it is greater in the rotationally dominated regime ($\Lambda < 1$) and reduced in the magnetically dominated regime ($\Lambda \geq 1$). Section \ref{sec3b1} showed that, in the rotationally dominated regime, penetration depth decreases with increasing rotation (i.e., decreasing $E$ from $10^{-3}$ to $10^{-5}$), while in the magnetically dominated regime, it remains largely unaffected by rotation. Section \ref{sec3b2} further indicated that penetration depth remains nearly invariant across diffusivity ratios ($q$), regardless of magnetic field strength. Based on these insights, we now quantify the penetration depth of $u_z^{\prime}$ by fixing $q = 1$ and varying $\Lambda$ from 0 to 10, for both weak ($h = 0.8$) and strong ($h = 0.6$) stable stratification, across three rotation rates ($E = 10^{-3}, 10^{-4}, 10^{-5}$). Herein, the penetration depth ($\delta_N$) of axial velocity perturbations ($u_z^{\prime}$) into the stably stratified layer above the interface ($z = h$) is defined as
\begin{equation} \label{eq_drop_u1}
\delta_{N} = z^{\dagger} - h,
\end{equation}
where $z^{\dagger}$ is the axial position at which the velocity decays to $1/e$ (approximately 37\%) of its value at the interface:
\begin{equation} \label{eq_drop_u2}
u_z^{\prime}|{z^{\dagger}} = \frac{1}{e} u_z^{\prime}|{z=h},
\end{equation}
with $e \approx 2.718$ being the base of the natural logarithm. This penetration depth, determined from direct numerical simulations, characterizes how far convective flows intrude into the stable layer. The corresponding percentage of penetration into the stable region ($z > h$) is calculated as
\begin{equation} \label{eq_drop_u3}
\delta_N (\%) = \frac{\delta_N}{(1 - h)} \times 100\%.
\end{equation}
This quantifies the fraction of the stable layer affected by upward convective motion from the unstable region below. 

\subsubsection{Variation with $\Lambda$}\label{sec3c1}
\begin{figure}[htbp]
    \centering
    \includegraphics[clip, trim=0cm 16cm 0cm 0cm, width=1\textwidth]{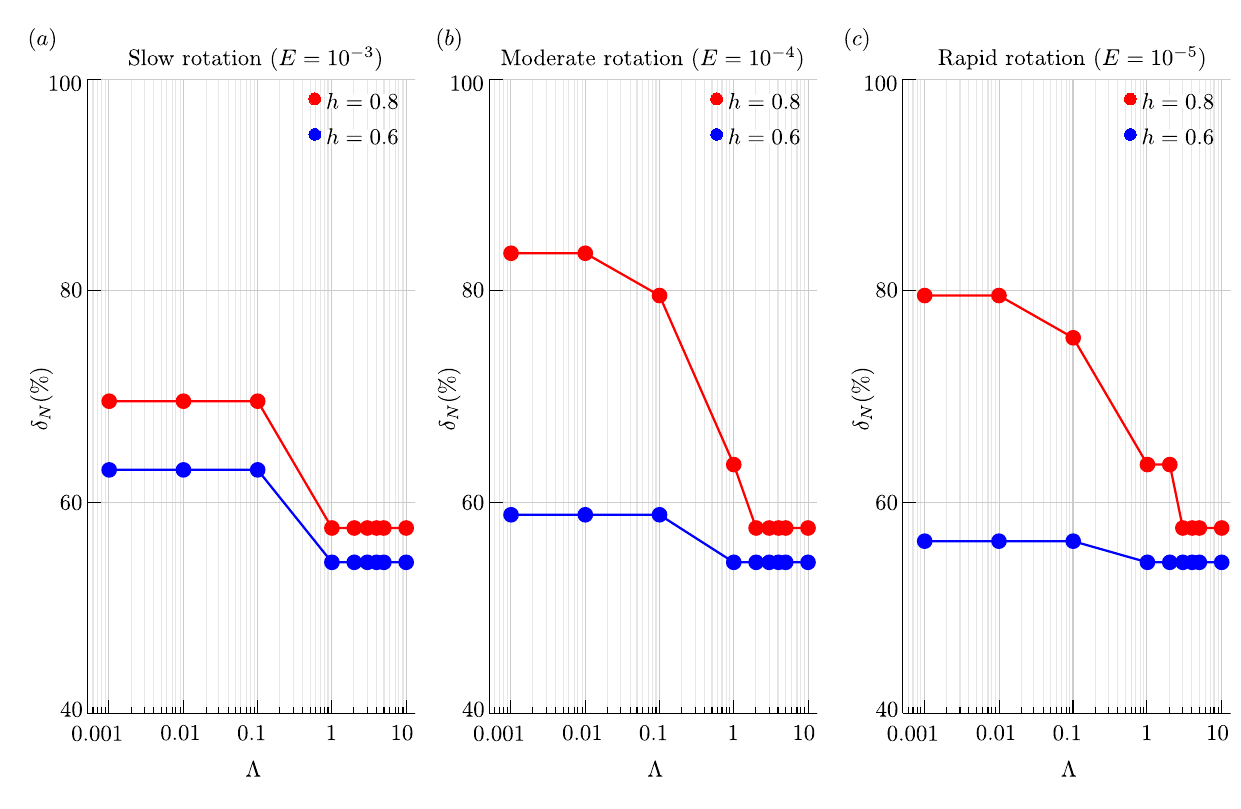}
    \caption{Percentage of penetration ($\delta_N(\%)$) vs. $\Lambda$ is plotted for moderate diffusivity ratio at $q = 1$ (a) for rapid rotation $E = 10^{-5}$, (b) for moderate rotation $E = 10^{-4}$, and (c) for slow rotation $E = 10^{-3}$. The weak ($h = 0.8$) and strong ($h = 0.6$) stable stratification cases are denoted by red and blue filled circles.}
    \label{f_penetration_depth}
\end{figure}

Regardless of rotation rates ($E$) and imposed magnetic field strength ($\Lambda$), the percentage penetration ($\delta_N (\%)$) of axial velocity ($u_z^{\prime}$) into stable stratification reduces as the strength of stable stratification increases for $h = 0.8$ (red circles) to $h = 0.6$ (blue circles) as depicted in figure (\ref{f_penetration_depth} a - c). Additionally, $\delta_N (\%)$ reduces as imposed magnetic field strength ($\Lambda$) increases, for both weak and strong imposed magnetic field regime, irrespective of rotation rates as shown in figure (\ref{f_penetration_depth}). It is attributed to stronger stabilization by Lorentz force as $\Lambda$ increases.

In the weak field regime ($\Lambda < 1$), the percentage penetration ($\delta_N(\%)$) decreases monotonically with increasing $\Lambda$ ($0.001 \rightarrow 0.1$) under rapid [Fig. \ref{f_penetration_depth}c] and moderate [Fig. \ref{f_penetration_depth}b] rotation for the weak stratification case ($h = 0.8$, red circles). For rapid rotation, $\delta_N(\%)$ remains around $\sim80\%$ for $\Lambda = 0.001, 0.01$, then drops to $\sim75\%$ at $\Lambda = 0.1$; similarly, for moderate rotation, it remains same to $\sim 84\%$ and drops to $\sim 80\%$ at $\Lambda = 0.1$. In contrast, for slow rotation ($E = 10^{-3}$), $\delta_N(\%)$ remains nearly constant at $\sim70\%$ for $\Lambda: 0.001 \rightarrow 0.1$ [Fig. \ref{f_penetration_depth}a]. For strong stratification ($h = 0.6$, blue circles), $\delta_N(\%)$ shows minimal dependence on $\Lambda$ ($0.001 \rightarrow 0.1$), but decreases with increasing rotation: from $\sim 63\%$ at $E = 10^{-3}$ [Fig. \ref{f_penetration_depth}a: blue circles] to $\sim 58\%$ at $E = 10^{-4}$ [Fig. \ref{f_penetration_depth}b: blue circles] and $\sim56\%$ at $E = 10^{-5}$ [Fig. \ref{f_penetration_depth}c: blue circles].

Transitioning from a weak ($\Lambda < 1$) to a strong magnetic field regime ($\Lambda \geq 1$) leads to a sharp decline in penetration, regardless of rotation rate or stratification strength, as shown in Figs. \ref{f_penetration_depth}a–c. For weak stratification ($h = 0.8$, red circles), under slow rotation ($E = 10^{-3}$), $\delta_N(\%)$ drops from $\sim 70\%$ at $\Lambda = 0.1$ to $\sim 57\%$ at $\Lambda = 1$, remaining constant for $\Lambda = 1 \rightarrow 10$ [Fig. \ref{f_penetration_depth}a]. At moderate rotation ($E = 10^{-4}$), it decreases to $\sim 63\%$ at $\Lambda = 1$ and further to $\sim 57\%$ for $\Lambda = 2 \rightarrow 10$ [Fig. \ref{f_penetration_depth}b]. For rapid rotation ($E = 10^{-5}$), $\delta_N(\%)$ holds at $\sim 63\%$ for $\Lambda = 1-2$, then drops to $\sim 57\%$ for $\Lambda = 3 \rightarrow 10$ [Fig. \ref{f_penetration_depth}c]. In the case of strong stratification ($h = 0.6$, blue circles), $\delta_N(\%)$ at $\Lambda = 0.1$ is $\sim 63\%$ at $E = 10^{-3}$, $\sim 58\%$ at $E = 10^{-4}$, and $\sim 56\%$ at $E = 10^{-5}$; all reduce to $\sim 54\%$ at $\Lambda = 1$ and remain nearly unchanged up to $\Lambda = 10$ across all rotation rates [Figs. \ref{f_penetration_depth}a–c]. This decline in penetration with increasing magnetic field strength is attributed to the suppressive influence of the Lorentz force in the strong field regime. 

The penetration effect is consistently higher for weak stratification ($h = 0.8$) than strong stratification ($h = 0.6$) across all $E$ and $\Lambda$ values [Fig. \ref{f_penetration_depth}], aligning with prior studies \cite{barman2025penetration, cai2020penetrative, xu2024penetrative}. In the weak field regime ($\Lambda: 0.001 \rightarrow 1$), penetration decreases monotonically with increasing rotation for strong stratification, but shows a non-monotonic trend for weak stratification—peaking at $E = 10^{-4}$ and lower at $E = 10^{-3}, 10^{-5}$ for $\Lambda: 0.001 \rightarrow 1$. This suggests optimal penetration at moderate rotation for weak stratification and at slow rotation for strong stratification. The contrasting influence of weak and strong stable stratification on penetration effect, based on rotation, is examined in greater detail next.

\subsubsection{Critical Ekman number ($E_c$)}\label{sec3c2}
Previous studies on penetrative thermal convection \cite{garai2022convective, barman2024role} showed enhanced penetration when transitioning from non-rotating to rotating states ($E = 10^{-3}$). In contrast, investigations of rapidly rotating thermal \cite{cai2020penetrative} and magnetoconvection \cite{xu2024penetrative} revealed reduced penetration as rotation intensified ($E \leq 10^{-4}$). In the previous section (\ref{sec3c1}), we observed that the maximum penetration occurs at different Ekman numbers ($E$) for weak and strong stratification, suggesting a critical Ekman number ($E_c$) where penetration peaks and diminishes on either side—likely influenced by stratification strength.

To isolate the pure rotational influence on the penetration effect and identify the critical Ekman number ($E_c$), we consider the non-magnetic case by neglecting magnetic effects. Specifically, we set $\Lambda = 0$ and fix $q = Pm = Pr = 1$, representing rotating thermal convection. The Ekman number ($E$) is varied from $10^{-6}$ to $10^{-1}$, including the non-rotating limit ($E = \infty$), to estimate the penetration depth for both weak ($h = 0.8$) and strong ($h = 0.6$) stratification. 

\begin{figure}[htbp!]
    \centering
    \includegraphics[clip, trim=0cm 21cm 0cm 0cm, width=1\textwidth]{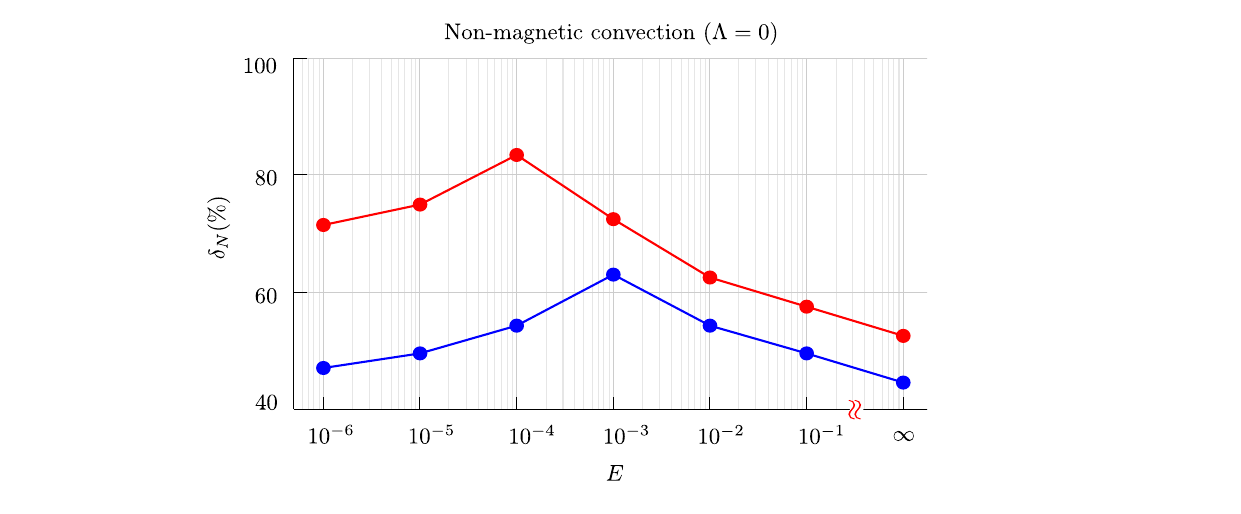}
    \caption{Percentage of penetration ($\delta_N(\%)$) vs. $E (10^{-6} \rightarrow \infty$) is plotted for non-magnetic case ($\Lambda = 0$). The weak and strong stable stratification cases are denoted by red and blue filled circles.}
    \label{f_critical_E} 
\end{figure}

As shown in figure (\ref{f_critical_E}), the percentage penetration is consistently higher for weaker stratification ($h = 0.8$, red dots) than for stronger stratification ($h = 0.6$, blue dots), since thicker stable layers more effectively suppress axial flow into the upper stable region. In both cases, penetration increases from the non-rotating limit to a maximum at a distinct Ekman number, then decreases on either side. For weak stable stratification model ($h = 0.8$), maximum penetration occurs at $E = 10^{-4}$, indicating the critical Ekman number ($E_c$). Whereas for strong stable stratification ($h = 0.6$), it peaks at $E_c = 10^{-3}$. This clearly demonstrates that the maximum penetration occurs at different critical Ekman numbers ($E_c$) depending on the strength of stable stratification. As stratification strengthens from $h = 0.8$ to $h = 0.6$, the peak penetration shifts toward slower rotation. However, the underlying mechanism behind this dichotomy is reserved for future investigation.

\section{Discussion}\label{sec4}
Previous studies on penetrative thermal convection in plane-layer models \cite{garai2022convective, barman2024role} have primarily investigated the effects of stratification in the absence of an imposed magnetic field. Further investigations on rotating magnetoconvection considered uniform magnetic fields applied in the axial \cite{chandrasekhar2013hydrodynamic} and horizontal \cite{roberts2000onset, jones2000onset} directions to analyze the characteristics of convective onset using linear stability analysis. Recent studies \cite{sahoo2023onset, sreenivasan2024oscillatory} have explored oscillatory magnetoconvection under spatially varying magnetic fields. Additionally, f-plane models \cite{cai2020penetrative, xu2024penetrative} have been employed to estimate penetration depths at various co-latitudes ($\theta$), using both wave and convection theories under uniform magnetic fields. The present study systematically investigates the onset of rotating magnetoconvection in a plane-layer model subject to a uniform axial magnetic field ($B^* = B_0 \mathbf{\hat{z}}$) and partial thermal stable stratification. We focus on the fundamental impacts of weak ($h = 0.8$) and strong ($h = 0.6$) stable layers on the onset characteristics, flow structures, symmetry breaking effects and penetration depth of convective instabilities. 

The onset of rotating magnetoconvection with an axially imposed magnetic field shows that oscillatory convection does not occur when $Pm < Pr$ (i.e., $q < 1$), as stated in equation (232) and the corresponding discussion in Chapter IV of Chandrasekhar (1961) \cite{chandrasekhar2013hydrodynamic}. The present study supports this finding: when magnetic diffusion dominates over thermal diffusion ($\kappa < \eta$), oscillatory modes are absent for $q < 1$ under fully unstable stratification. Moreover, when magnetic and thermal diffusion are comparable ($q = 1$), oscillatory modes remain absent across $\Lambda = 0 \rightarrow 10$. Introducing stable stratification—whether weak or strong—does not alter this behavior for low and moderate diffusivity ratios ($q = 0.01, 1$) across slow to rapid rotation rates. For high diffusivity ratios ($q = 10$), the situation changes: oscillatory convection appears in both weak- and strong-field regimes, but at $\Lambda = 1$ the mode becomes stationary, marking a transition from a viscous oscillatory mode for $\Lambda: 0 \rightarrow 0.1$ to a magnetic oscillatory mode for $\Lambda: 2 \rightarrow 10$. This trend persists across all rotation rates and stratification models. A detailed quantitative analysis of oscillatory mode characteristics and the critical magnetic field strength ($\Lambda_c$) marking the steady–oscillatory transition is reserved for future work.  


In horizontal magnetoconvection (HMC) \cite{barman2025penetration}, convective rolls thicken monotonically with increasing horizontal magnetic field strength ($\Lambda$) due to axial velocity stretching by the imposed field, $(\mathbf{B}^* \cdot \nabla)\mathbf{u}$. In the present study with an axially imposed field, fully unstable stratification without a magnetic field ($\Lambda = 0$) produces geostrophic columns, whereas increasing $\Lambda > 0$ modifies roll structure non-monotonically by relaxing rotational constraints and promoting quasi-2D large-scale organization. At $\Lambda = 1$, rolls become thickest due to the combined effects of rotational relaxation and magnetic stretching. Under a strongly imposed axial field ($\Lambda > 1$), axial flow is stretched vertically, producing elongated rolls; at high $\Lambda$, this leads to partial restoration of columnarity and increased wavenumber, reflecting magnetic dominance in setting convective scales. This trend persists even in the presence of stable stratification. 

In penetrative thermal convection \cite{garai2022convective, barman2024role}, the stable stratification confines temperature perturbations entirely within the unstable layer, while axial velocity can penetrate the stable layer. Similar behavior is observed in the weak-field regime ($\Lambda < 1$); however, at $\Lambda = 1$, temperature perturbations propagate into the stable layer for both weak and strong stratification across all $E$. A further increase to $\Lambda = 5$ again suppresses temperature perturbations within the unstable layer. This non-monotonic suppression with $\Lambda$ may arise from magnetic back-reaction effects under strong imposed fields. In contrast, in horizontally imposed magnetoconvection, suppression of temperature perturbations decreases monotonically with increasing $\Lambda$ from $1 \rightarrow 10$ for both weak and strong stratification, as found in previous investigation\cite{barman2025penetration}. The lack of a clear correlation between temperature and magnetic field perturbations indicates that further theoretical and numerical investigation is required. 

Fully unstable stratification exhibits perfect symmetry in convective instabilities. Stable stratification, however, breaks this axial symmetry-a hallmark of penetrative convection \cite{garai2022convective, barman2025penetration}. As stable stratification strengthens, from $h = 0.8$ to $h = 0.6$, asymmetry increases in both axial velocity and temperature perturbations. In the weak-field regime, this symmetry-breaking effect is stronger than in the strong-field regime, regardless of stratification strength, indicating that a strong magnetic field tends to restore symmetry and mitigate the influence of stable stratification. Further, rapidly rotating magnetoconvection \cite{aujogue2015onset} under fully unstable stratification exhibits global scalings, with $Ra_c, k_c^x \propto E^{-1/3}$ in weak fields and $Ra_c \propto \Lambda^{-1}, k_c^x$ independent of $E$ and $\Lambda$, in strong fields. Building on this, we analyze convection onset with an axially imposed magnetic field in the presence of stable stratification. Instead of global relations, we derive local scalings for rotation-dominated ($\Lambda < 1$) and magnetically dominated ($\Lambda \geq 1$) regimes across three stratification models, rotation rates ($E = 10^{-3}, 10^{-4}, 10^{-5}$), and diffusivity ratios ($q = 0.01, 1, 10$). We find that in weak fields $Ra_c$ and $k_c^x$ depend only weakly on $\Lambda$ ($\propto \Lambda^{-0.01}$), while in strong fields they increase for $q = 0.01, 1$ but decrease for $q = 10$, highlighting reduced magnetic stabilization when thermal diffusivity dominates.

Penetrative convection influences geomagnetic secular variation and the geodynamo \cite{takehiro2001penetration, aubert2025rapid}, making accurate estimation of penetration depth crucial for understanding stable layer dynamics. This depth depends on both rotation rate ($E$) and magnetic field strength ($\Lambda$). Previous studies \cite{garai2022convective, barman2025penetration} found that in slowly rotating regimes ($E \geq 10^{-3}$), rotation enhances penetration, whereas under rapid rotation ($E \leq 10^{-3}$) it decreases with increasing rotation—peaking at $E = 10^{-4}$ for weak stratification \cite{cai2020penetrative}. The present study corroborates this for weak stratification, but reveals that for strong stratification, maximum penetration occurs at $E = 10^{-3}$. Additionally, a background magnetic field stabilizes the system and reduces penetration, consistent with previous studies\cite{xu2024penetrative}.

\section{Conclusions}\label{sec5}
In this work, we investigate the influence of partial stable stratification on the onset of rotating magnetoconvection under a uniform vertical magnetic field. A detailed parametric study is conducted for $E = 10^{-3}, 10^{-4}, 10^{-5}$, $q = 0.01, 1, 10$, $\Lambda = 0$–$10$, and $h = \infty, 0.8, 0.6$, leading to the following key outcomes:

\begin{itemize}
    \item Stable stratification can break the axial symmetry of convective flows, influencing dynamo action and geomagnetic secular variation. This symmetry-breaking is most pronounced in the weak-field regime but reduces under strong fields, indicating suppression of stable layer impact by the Lorentz force. Stable stratification also lowers the critical Rayleigh number ($Ra_c$) compared to fully unstable cases. As the system shifts from fully unstable to weakly and strongly stably stratified states, the critical horizontal wavenumber increases in both rotation- and magnetically dominated regimes. The stabilizing influence of stable stratification is stronger in rotation-dominated regimes ($\Lambda < 1$) than in magnetically dominated regimes ($\Lambda \geq 1$). 
    
    \item In weak field regime ($\Lambda < 1$), columnar rolls appear due to rotational constraint. At $\Lambda  = 1$, the columnarity breaks down resulting in thickest convective rolls. The columnarity restores as system is introduced to stronger background magnetic field ($\Lambda > 1$) due to elongation of axial velocity in the direction of axially imposed magnetic field. The thickening of convection rolls with $\Lambda$ occur non-monotonically as $\Lambda$ increases. For low to moderate diffusivity ratios ($q = 0.01, 1$), magnetic stabilization becomes more effective in raising the onset threshold with $\Lambda$, whereas this effect weakens at high diffusivity ($q = 10$) across all stratification models.
    
    \item The percentage penetration ($\delta_N(\%)$) decreases with increasing magnetic field strength, as the Lorentz force stabilizes the system and suppresses penetration. In the weak-field regime ($\Lambda < 1$), $\delta_N(\%)$ decreases monotonically with rotation for strong stratification ($h = 0.6$) but exhibits non-monotonic variation for weak stratification ($h = 0.8$). In the magnetically dominated regime ($\Lambda \geq 1$), the penetration depth remains nearly unchanged with rotation across all stratification cases.
    
    \item In the non-magnetic case ($\Lambda = 0$), the maximum percentage penetration ($\delta_N(\%)$) for weak and strong stable stratification occurs at different Ekman numbers ($E$), revealing a critical Ekman number ($E_c$) for maximum penetration depth. For weak stratification ($h = 0.8$) and strong stratification ($h = 0.6$), $E_c$ is $10^{-4}$ and $10^{-3}$, respectively, with penetration decreasing on either side of $E_c$.  
\end{itemize}

The present study examines the role of partial thermal stable stratification in controlling the onset of magnetoconvection under a uniform axial magnetic field, spanning a wide range of diffusivity ratios and rotation rates. The results hold strong relevance for geophysical and astrophysical fluid dynamics, particularly in planetary cores and the solar interior. While key aspects across broad parameter regimes have been addressed, further work is required to understand how stable layers influence nonlinear dynamics under strong buoyancy forcing.

\section*{ACKNOWLEDGMENTS}
Authors would like to thanks Mr. Arpan Das for fruitful discussion. S.S. acknowledges the financial support through the award of the INSPIRE Faculty Fellowship by the Department of Science and Technology, India (Grant No. IFA18-EAS 70).

\bibliographystyle{acm}
\bibliography{reference}

\begin{thebibliography}{10}

\bibitem{aubert2025rapid}
{\sc Aubert, J.}
\newblock Rapid geomagnetic variations and stable stratification at the top of
  earth's core.
\newblock {\em Physics of the Earth and Planetary Interiors 362\/} (2025),
  107335.

\bibitem{aujogue2015onset}
{\sc Aujogue, K., Poth{\'e}rat, A., and Sreenivasan, B.}
\newblock Onset of plane layer magnetoconvection at low ekman number.
\newblock {\em Physics of Fluids 27}, 10 (2015).

\bibitem{barman2024role}
{\sc Barman, T., and Sahoo, S.}
\newblock Role of partial stable stratification on fluid flow and heat transfer
  in rotating thermal convection.
\newblock {\em Physics of Fluids 36}, 4 (2024).

\bibitem{barman2025penetration}
{\sc Barman, T., and Sahoo, S.}
\newblock Penetration depth of flow perturbations into a stably stratified
  layer driven by rotating laterally heterogeneous convection.
\newblock {\em Physics of Fluids 37}, 8 (2025).

\bibitem{brodholt2017composition}
{\sc Brodholt, J., and Badro, J.}
\newblock Composition of the low seismic velocity e' layer at the top of
  earth's core.
\newblock {\em Geophysical Research Letters 44}, 16 (2017), 8303--8310.

\bibitem{brun2005simulations}
{\sc Brun, A.~S., Browning, M.~K., and Toomre, J.}
\newblock Simulations of core convection in rotating a-type stars: magnetic
  dynamo action.
\newblock {\em The Astrophysical Journal 629}, 1 (2005), 461.

\bibitem{brun2017differential}
{\sc Brun, A.~S., Strugarek, A., Varela, J., Matt, S.~P., Augustson, K.~C.,
  Emeriau, C., DoCao, O.~L., Brown, B., and Toomre, J.}
\newblock On differential rotation and overshooting in solar-like stars.
\newblock {\em The Astrophysical Journal 836}, 2 (2017), 192.

\bibitem{buffett2014geomagnetic}
{\sc Buffett, B.}
\newblock Geomagnetic fluctuations reveal stable stratification at the top of
  the earth’s core.
\newblock {\em Nature 507}, 7493 (2014), 484--487.

\bibitem{buffett2016evidence}
{\sc Buffett, B., Knezek, N., and Holme, R.}
\newblock Evidence for mac waves at the top of earth's core and implications
  for variations in length of day.
\newblock {\em Geophysical Journal International 204}, 3 (2016), 1789--1800.

\bibitem{buffett2010stratification}
{\sc Buffett, B.~A., and Seagle, C.~T.}
\newblock Stratification of the top of the core due to chemical interactions
  with the mantle.
\newblock {\em Journal of Geophysical Research: Solid Earth 115}, B4 (2010).

\bibitem{burns2019dedalus}
{\sc Burns, K., Vasil, G., Oishi, J., Lecoanet, D., and Brown, B.}
\newblock Dedalus: A flexible framework for numerical simulations with spectral
  methods. arxiv e-prints.
\newblock {\em arXiv preprint arXiv:1905.10388\/} (2019).

\bibitem{cai2020penetrative}
{\sc Cai, T.}
\newblock Penetrative convection for rotating boussinesq flow in tilted
  f-planes.
\newblock {\em The Astrophysical Journal 898}, 1 (2020), 22.

\bibitem{calkins2023numerical}
{\sc Calkins, M.~A., AlRefae, T., Hernandez, A., Yan, M., and Maffei, S.}
\newblock Numerical investigation of quasistatic magnetoconvection with an
  imposed horizontal magnetic field.
\newblock {\em Physical Review Fluids 8}, 12 (2023), 123501.

\bibitem{cattaneo2006dynamo}
{\sc Cattaneo, F., and Hughes, D.~W.}
\newblock Dynamo action in a rotating convective layer.
\newblock {\em Journal of Fluid Mechanics 553\/} (2006), 401--418.

\bibitem{chandrasekhar2013hydrodynamic}
{\sc Chandrasekhar, S.}
\newblock {\em Hydrodynamic and hydromagnetic stability}.
\newblock Courier Corporation, 2013.

\bibitem{dietrich2018penetrative}
{\sc Dietrich, W., and Wicht, J.}
\newblock Penetrative convection in partly stratified rapidly rotating
  spherical shells.
\newblock {\em Frontiers in Earth Science 6\/} (2018), 189.

\bibitem{fortney2023saturn}
{\sc Fortney, J., Militzer, B., Mankovich, C., Helled, R., Wahl, S.,
  Nettelmann, N., Hubbard, W., Stevenson, D., Iess, L., Marley, M., et~al.}
\newblock Saturn's interior after the cassini grand finale.
\newblock {\em arXiv preprint arXiv:2304.09215\/} (2023).

\bibitem{garai2022convective}
{\sc Garai, S., and Sahoo, S.}
\newblock On convective instabilities in a rotating fluid with stably
  stratified layer and thermally heterogeneous boundary.
\newblock {\em Physics of Fluids 34}, 12 (2022).

\bibitem{gastine2020dynamo}
{\sc Gastine, T., Aubert, J., and Fournier, A.}
\newblock Dynamo-based limit to the extent of a stable layer atop earth’s
  core.
\newblock {\em Geophysical Journal International 222}, 2 (2020), 1433--1448.

\bibitem{glatzmaier2013introduction}
{\sc Glatzmaier, G.~A.}
\newblock Introduction to modeling convection in planets and stars: Magnetic
  field, density stratification, rotation.
\newblock In {\em Introduction to Modeling Convection in Planets and Stars}.
  Princeton University Press, 2013.

\bibitem{gubbins2007geomagnetic}
{\sc Gubbins, D.}
\newblock Geomagnetic constraints on stratification at the top of earth’s
  core.
\newblock {\em Earth, planets and space 59\/} (2007), 661--664.

\bibitem{gubbins2013stratified}
{\sc Gubbins, D., and Davies, C.}
\newblock The stratified layer at the core--mantle boundary caused by
  barodiffusion of oxygen, sulphur and silicon.
\newblock {\em Physics of the Earth and Planetary Interiors 215\/} (2013),
  21--28.

\bibitem{helled2024fuzzy}
{\sc Helled, R., and Stevenson, D.~J.}
\newblock The fuzzy cores of jupiter and saturn.
\newblock {\em AGU Advances 5}, 2 (2024), e2024AV001171.

\bibitem{hurlburt1994penetration}
{\sc Hurlburt, N.~E., Toomre, J., Massaguer, J.~M., and Zahn, J.-P.}
\newblock Penetration below a convective zone.
\newblock {\em Astrophysical Journal, Part 1 (ISSN 0004-637X), vol. 421, no. 1,
  p. 245-260 421\/} (1994), 245--260.

\bibitem{jones2000onset}
{\sc Jones, C., and Roberts, P.}
\newblock The onset of magnetoconvection at large prandtl number in a rotating
  layer ii. small magnetic diffusion.
\newblock {\em Geophysical \& Astrophysical Fluid Dynamics 93}, 3-4 (2000),
  173--226.

\bibitem{jones2000convection}
{\sc Jones, C.~A., and Roberts, P.~H.}
\newblock Convection-driven dynamos in a rotating plane layer.
\newblock {\em Journal of Fluid Mechanics 404\/} (2000), 311--343.

\bibitem{kaneshima2018array}
{\sc Kaneshima, S.}
\newblock Array analyses of smks waves and the stratification of earth’s
  outermost core.
\newblock {\em Physics of the Earth and Planetary Interiors 276\/} (2018),
  234--246.

\bibitem{kundu2024fluid}
{\sc Kundu, P.~K., Cohen, I.~M., Dowling, D.~R., and Capecelatro, J.}
\newblock {\em Fluid mechanics}.
\newblock Elsevier, 2024.

\bibitem{lister2004thermal}
{\sc Lister, J.~R.}
\newblock Thermal winds forced by inhomogeneous boundary conditions in
  rotating, stratified, hydromagnetic fluid.
\newblock {\em Journal of Fluid Mechanics 505\/} (2004), 163--178.

\bibitem{lister1998stratification}
{\sc Lister, J.~R., and Buffett, B.~A.}
\newblock Stratification of the outer core at the core-mantle boundary.
\newblock {\em Physics of the earth and planetary interiors 105}, 1-2 (1998),
  5--19.

\bibitem{masada2013effects}
{\sc Masada, Y., Yamada, K., and Kageyama, A.}
\newblock Effects of penetrative convection on solar dynamo.
\newblock {\em The Astrophysical Journal 778}, 1 (2013), 11.

\bibitem{mason2022magnetoconvection}
{\sc Mason, S.~J., Guervilly, C., and Sarson, G.~R.}
\newblock Magnetoconvection in a rotating spherical shell in the presence of a
  uniform axial magnetic field.
\newblock {\em Geophysical \& Astrophysical Fluid Dynamics 116}, 5-6 (2022),
  458--498.

\bibitem{matsumoto2009does}
{\sc Matsumoto, M.}
\newblock Why does water expand when it cools?
\newblock {\em Physical review letters 103}, 1 (2009), 017801.

\bibitem{moore2022dynamo}
{\sc Moore, K., Barik, A., Stanley, S., Stevenson, D., Nettelmann, N., Helled,
  R., Guillot, T., Militzer, B., and Bolton, S.}
\newblock Dynamo simulations of jupiter's magnetic field: The role of stable
  stratification and a dilute core.
\newblock {\em Journal of Geophysical Research: Planets 127}, 11 (2022),
  e2022JE007479.

\bibitem{mukherjee2023thermal}
{\sc Mukherjee, P., and Sahoo, S.}
\newblock Thermal convection and dynamo action with stable stratification at
  the top of the earth's outer core.
\newblock {\em Physics of the Earth and Planetary Interiors 345\/} (2023),
  107111.

\bibitem{nicoski2022quasistatic}
{\sc Nicoski, J.~A., Yan, M., and Calkins, M.~A.}
\newblock Quasistatic magnetoconvection with a tilted magnetic field.
\newblock {\em Physical Review Fluids 7}, 4 (2022), 043504.

\bibitem{olson1995magnetoconvection}
{\sc Olson, P., and Glatzmaier, G.~A.}
\newblock Magnetoconvection in a rotating spherical shell: structure of flow in
  the outer core.
\newblock {\em Physics of the Earth and Planetary Interiors 92}, 1-2 (1995),
  109--118.

\bibitem{olson1996magnetoconvection}
{\sc Olson, P., and Glatzmaier, G.~A.}
\newblock Magnetoconvection and thermal coupling of the earth's core and
  mantle.
\newblock {\em Philosophical Transactions of the Royal Society of London.
  Series A: Mathematical, Physical and Engineering Sciences 354}, 1711 (1996),
  1413--1424.

\bibitem{olson2010treatise}
{\sc Olson, P.~L.}
\newblock Treatise on geophysics, volume 8: Core dynamics.

\bibitem{peckover1978dynamic}
{\sc Peckover, R., and Weiss, N.}
\newblock On the dynamic interaction between magnetic fields and convection.
\newblock {\em Monthly Notices of the Royal Astronomical Society 182}, 2
  (1978), 189--208.

\bibitem{proctor1982magnetoconvection}
{\sc Proctor, M., and Weiss, N.}
\newblock Magnetoconvection.
\newblock {\em Reports on Progress in Physics 45}, 11 (1982), 1317.

\bibitem{roberts2000onset}
{\sc Roberts, P., and Jones, C.}
\newblock The onset of magnetoconvection at large prandtl number in a rotating
  layer i. finite magnetic diffusion.
\newblock {\em Geophysical \& Astrophysical Fluid Dynamics 92}, 3-4 (2000),
  289--325.

\bibitem{sahoo2023onset}
{\sc Sahoo, S., and ES, S.}
\newblock Onset of oscillatory magnetoconvection under rapid rotation and
  spatially varying magnetic field.
\newblock {\em Physics of Fluids 35}, 2 (2023).

\bibitem{sakuraba2002linear}
{\sc Sakuraba, A.}
\newblock Linear magnetoconvection in rotating fluid spheres permeated by a
  uniform axial magnetic field.
\newblock {\em Geophysical \& Astrophysical Fluid Dynamics 96}, 4 (2002),
  291--318.

\bibitem{schubert2000dynamics}
{\sc Schubert, G., and Zhang, K.}
\newblock Dynamics of giant planet interiors.
\newblock In {\em From Giant Planets to Cool Stars\/} (2000), vol.~212, p.~210.

\bibitem{sreenivasan2017confinement}
{\sc Sreenivasan, B., and Gopinath, V.}
\newblock Confinement of rotating convection by a laterally varying magnetic
  field.
\newblock {\em Journal of Fluid Mechanics 822\/} (2017), 590--616.

\bibitem{sreenivasan2024oscillatory}
{\sc Sreenivasan, S.~E., and Sahoo, S.}
\newblock Oscillatory onset of rotating thermal convection subject to spatially
  varying magnetic fields and stable stratification.
\newblock {\em Physics of Fluids 36}, 8 (2024).

\bibitem{stevenson2020jupiter}
{\sc Stevenson, D.~J.}
\newblock Jupiter's interior as revealed by juno.
\newblock {\em Annual Review of Earth and Planetary Sciences 48}, 1 (2020),
  465--489.

\bibitem{takehiro2001penetration}
{\sc Takehiro, S.-i., and Lister, J.~R.}
\newblock Penetration of columnar convection into an outer stably stratified
  layer in rapidly rotating spherical fluid shells.
\newblock {\em Earth and Planetary Science Letters 187}, 3-4 (2001), 357--366.

\bibitem{thelen2000dynamo}
{\sc Thelen, J.-C., and Cattaneo, F.}
\newblock Dynamo action driven by convection: the influence of magnetic
  boundary conditions.
\newblock {\em Monthly Notices of the Royal Astronomical Society 315}, 2
  (2000), L13--L17.

\bibitem{veronis1963penetrative}
{\sc Veronis, G.}
\newblock Penetrative convection.
\newblock {\em Astrophysical Journal, vol. 137, p. 641 137\/} (1963), 641.

\bibitem{xu2024penetrative}
{\sc Xu, F., and Cai, T.}
\newblock Penetrative magneto-convection of a rotating boussinesq flow in
  f-planes.
\newblock {\em Physics of Fluids 36}, 2 (2024).

\bibitem{yan2021recipe}
{\sc Yan, C., and Stanley, S.}
\newblock Recipe for a saturn-like dynamo.
\newblock {\em AGU Advances 2}, 2 (2021), e2020AV000318.

\bibitem{yan2019heat}
{\sc Yan, M., Calkins, M.~A., Maffei, S., Julien, K., Tobias, S.~M., and Marti,
  P.}
\newblock Heat transfer and flow regimes in quasi-static magnetoconvection with
  a vertical magnetic field.
\newblock {\em Journal of Fluid Mechanics 877\/} (2019), 1186--1206.

\bibitem{zahn1991convective}
{\sc Zahn, J.-P.}
\newblock Convective penetration in stellar interiors.
\newblock {\em Astronomy and Astrophysics (ISSN 0004-6361), vol. 252, no. 1,
  Dec. 1991, p. 179-188. 252\/} (1991), 179--188.

\bibitem{zhang2002penetrative}
{\sc Zhang, K., and Schubert, G.}
\newblock From penetrative convection to teleconvection.
\newblock {\em The Astrophysical Journal 572}, 1 (2002), 461.

\end{thebibliography}

\end{document}